\documentclass[pra,twocolumn,superscriptaddress,longbibliography]{revtex4-1}
\usepackage{amssymb,amsmath,amsfonts,bm}
\usepackage{graphicx}
\usepackage{epstopdf}
\usepackage{times}
\usepackage{float}
\usepackage{lipsum}
\usepackage{color}
\usepackage{ulem}

\usepackage{hyperref}
\hypersetup{colorlinks=true, linkcolor=blue, citecolor=red, urlcolor=blue  }

\usepackage{physics}

\begin{document}

\title{Error-Resilient Floquet Geometric Quantum Computation}%

\author{Yuan-Sheng Wang}
 \affiliation{Department of Physics, Southern University of Science and Technology, Shenzhen 518055, China}
 \affiliation{School of Physical Sciences, University of Science and Technology of China, Hefei 230026, China}

\author{Bao-Jie Liu}
 \affiliation{Department of Physics, Southern University of Science and Technology, Shenzhen 518055, China}
 
\author{Shi-Lei Su} \affiliation{School of Physics, Zhengzhou University, Zhengzhou 450001, China}

\author{Man-Hong Yung}  \email{yung@sustech.edu.cn}
\affiliation{Department of Physics, Southern University of Science and Technology, Shenzhen 518055, China}
\affiliation{Shenzhen Institute for Quantum Science and Engineering, Southern University of Science and Technology, Shenzhen 518055, China}
\affiliation{Guangdong Provincial Key Laboratory of Quantum Science and Engineering, Southern University of Science and Technology, Shenzhen 518055, China}
\affiliation{Shenzhen Key Laboratory of Quantum Science and Engineering, Southern University of Science and Technology, Shenzhen,518055, China}

\date{\today}

\begin{abstract}

We proposed a new geometric quantum computation (GQC) scheme, called Floquet GQC (FGQC), where error-resilient geometric gates based on periodically driven two-level systems can be constructed via a new non-Abelian geometric phase proposed in a recent study [V. Novi\^{c}enko \textit{et al}, Phys. Rev. A 100, 012127 (2019) ]. 
Based on Rydberg atoms, we gave possible implementations of universal single-qubit gates and a nontrivial two-qubit gate for FGQC.
By using numerical simulation, we evaluated the performance of the FGQC Z and X gates in the presence of both decoherence and a certain kind of systematic control error. 
The gate fidelities of the Z and X gates are $F_{X,\text{gate}}\approx F_{Z,\text{gate}}\approx 0.9992$.
The numerical results provide evidences that FGQC gates can achieve fairly high gate fidelities even in the presence of noise and control imperfection.
In addition, we found FGQC is robust against global control error, both analytical demonstration and numerical evidence were given.
Consequently, this study makes an important step towards robust geometric quantum computation. 

\end{abstract}

\maketitle


\section{Introduction}

Quantum computations can solveecertain problems much more effectively than classical computations, such as quantum simulations \cite{feynman1999,PhysRevLett.112.220501}, prime factoring \cite{Vandersypen2001,Xu2012,Martin-Lopez2012}, searching unsorted data \cite{Grover1997} and machine learning \cite{Rebentrost2014,PhysRevLett.114.140504,Cong2019}.
High-fidelity universal gates for quantum bits form an essential ingredient of quantum computation and quantum information processing.
One promising approach towards this goal is to use geometric phases~\cite{berry1984quantal,Aharonov1987,Wilczek1984,ANANDAN1988171} that have attracted much attentions because geometric phases depend only on the global properties of the evolution trajectories, and thus have built-in noise-resilient features against certain local noises~\cite{Chiara2003Berry,2005Geometric,Leek1889,2009Experimental,2013Exploring}. 

The early applications of geometric quantum computation (GQC) depend on adiabatic quantum evolution to suppress transitions between different instantaneous eigenstates of Hamiltonian \cite{Jones2000,Duan1695,Wu2005,Wu2013,Huang2019}.
These adiabatic gates operate slowly compared to the dynamical time scale. They become vulnerable to decoherence effects that may lead to the loss of coherence.
To overcome the dilemma between the limited coherence times and the long duration of adiabatic evolution, nonadiabatic geometric quantum computation (NGQC)  \cite{Xiang-Bin2001,Zhu2002,Thomas2011,Zhao2017,Li2020,Chen2018,Zhang2020} and nonadiabatic holonomic quantum computation (NHQC)  \cite{Liu2019,Sjoeqvist2012,Xugf2012,Xue2015,Xue2017,Zhou2017,Hong2018,AzimiMousolou2017,Zhao2020,Johansson2012,Zheng2016,Ramberg2019,Jing2017,Liu2020} based on nonadiabatic Abelian and non-Abelian geometric gate \cite{Aharonov1987,ANANDAN1988171} have been proposed.
Owing to its intrinsic noise-resistance features and high-speed in implementation, NGQC has attracted considerable interest \cite{Zhao2017,Xue2015,Feng2013,Li2017,AbdumalikovJr2013,Xu2018,Zhang2019,Danilin2018,Egger2019,Zu2014,Arroyo-Camejo2014,Zhou2017a,Sekiguchi2017,Nagata2018,Ishida:18,Zhu2003,Feng2009,Ota2009,Spiegelberg2013,Xu2014,Liang2014,Xu2015,Albert2016,Zhao2016,Zhao2017a,Zhao2018,Zhao2019,Leibfried2003,Du2006,Liu2019} and has been experimentally demonstrated with nuclear magnetic resonance \cite{Feng2013,Li2017}, superconducting circuits \cite{AbdumalikovJr2013,Xu2018,Zhang2019,Danilin2018,Egger2019,Yan2019,Xu2020,zhao2019experimental} and nitrogen-vacancy centers in diamond \cite{Zu2014,Arroyo-Camejo2014,Zhou2017a,Sekiguchi2017,Nagata2018,Ishida:18}. However, these nonadiabatic gates require the driving pulses to satisfy the restrictive conditions, thus reducing the robustness of the resulting geometric gates against control errors~\cite{Zheng2016,Ramberg2019,Jing2017,Liu2020}. 
On the other hand, the use of periodic control pulses has emerged as a ubiquitous tool for the coherent control and engineering of quantum dynamics\cite{Haeberlen1968,Vandersypen2005,Goldman2014,Bukov2015,Poudel2015,Eckardt2017,Oka2019}, with the recent increase in applications in constructing GQC scenarios\cite{Novifmmodeheckclsecienko2017,PhysRevA.100.012127,Bomantara2018,Bomantara2018a}.
    
In this paper, we proposed a new geometric computation scheme, called Floquet GQC (FGQC), where universal error-resistant geometric gates can be constructed via a new non-Abelian geometric phase. 
This non-Abelian geometric phase emerges from a periodically driven quantum systems and was found in a recent study~\cite{PhysRevA.100.012127}. 
Based on Rydberg atoms, a possible experimental realization of our proposal was provided.   
By applying this proposal, in the presence of decoherence and a certain kind of systematic control error, numerical simulations of the Z and X gates were given. 
The relevant results are shown in Figure \ref{fig:pulses}, which indicate that even in the presence of reasonable degree of decoherence and error, FGQC gates can still achieve fairly high fidelities.  
Furthermore, we found FGQC is robust against global control error. 
A analytical demonstration was given.  
In addition, using the a possible set of parameters, we simulated the FGQC X and Z gates in the presence of global control error, then we compared their performances with that of the same gates based on DG and NGQC scenarios ~\cite{Xiang-Bin2001,Zhu2002,Thomas2011,Zhao2017,Li2020,Chen2018,Zhang2020}, the results are shown in Fig. \ref{fidelity}. 
These numerical results provide evidence of the superiority of FGQC gates over NGQC and standard dynamical gate (DG) gates, in terms of solving global control error.

This paper is organized as follows. In Sec. \ref{sec:general}, the general theory of GQC and FGQC is introduced. 
In Sec. \ref{sec:physics}, a possible implementation of universal single-qubit and a nontrivial  two-qubit gate using Rydberg atoms is extensively studied concretely based on the proposed theory in Sec. \ref{sec:general}. 
Sec. \ref{sec:conclusion} summarizes the findings.

\section{\label{sec:general}General theory}
\subsection{General theory of GQC}

Consider a quantum system exposed to Hamiltonian $H(t)$.
For any set of complete basis vectors $\{|\psi_{\alpha}(0)\rangle\}$ at $t=0$, the unitary temporal evolution operator can be expressed as $U(t,0)=\mathcal{T}e^{-i\int_{0}^{t^{\prime}}H(t^{\prime})dt^{\prime}}=\sum_{\alpha}|\psi_{\alpha}(t)\rangle\langle\psi_{\alpha}(0)|$, where $\mathcal{T}$ is the time-ordering operator, $|\psi_{\alpha}(t)\rangle=\mathcal{T}e^{-i\int_{0}^{t^{\prime}}H(t^{\prime})dt^{\prime}}|\psi_{\alpha}(0)\rangle$ is a time-dependent state satisfying the Schr\"odinger equation.
Now, at each moment of time, a different set of time-dependent basis can be selected, $\{|\mu_{\alpha}(t)\rangle\}$, 
satisfying the boundary conditions at $t=0$ and $t=\tau$:
\begin{align}
    |\mu_{\alpha}(\tau)\rangle =|\mu_{\alpha}(0)\rangle=|\psi_{\alpha}(0)\rangle.
    \label{c_condition}
\end{align}
The time evolution state can be written as $    |\psi_{\alpha}(t)\rangle=\sum_{\beta}c_{\alpha\beta}(t)|\mu_{\beta}(t)\rangle$. 
By substituting this equation into the Schr\"odinger equation, we have  
\begin{align}
 \frac{d}{dt}{c}_{\alpha \beta}(t)=i\sum_{\gamma}[A_{\alpha\gamma}(t)-H_{\alpha\gamma}(t)]c_{\gamma \beta}(t),
 \label{derivative}
\end{align}
where $H_{\alpha\beta}(t)=\langle\mu_{\alpha}(t)|H(t)|\mu_{\beta}(t)\rangle$ and $A_{\alpha\beta}(t)=\langle\mu_{\alpha}(t)|i\partial_{t}|\mu_{\beta}(t)\rangle|$, 
which can be combined to form an effective Hamiltonian:
$H_{\text{eff}}(t)=R^{\dagger}(t)[H(t)-id/dt]R(t)$, with $R(t)=\sum_{\alpha}|\mu_{\alpha}(t)\rangle\langle\mu_{\alpha}(0)|$.
Consider any unitary once differentiable operator $V(t)$ satisfying the boundary condition $V(\tau)=V(0)$.
The operator $A(t)$ is transformed as a proper gauge potential under the basis change: 
$|\mu_{\alpha}(t)\rangle\rightarrow \sum_{\gamma}|\mu_{\alpha}(t)\rangle V_{\alpha\gamma}(t)$, 
and $A(t)$ will generally contribute a non-Abelian geometric phase to the temporal evolution operator $U(\tau,0)$. 

For a GQC scenario, the dynamical phase should be eliminated by engineering $H(t)$ and the ancillary basis $\{|\mu_{\alpha}(t)\rangle\}$. 
If the effective Hamiltonian is always diagonal in the initial basis, $A_{\alpha\beta}(t)-H_{\alpha\beta}(t)=\delta_{\alpha\beta}[A_{\alpha\alpha}(t)-H_{\alpha\alpha}(t)]$. 
Then, the time evolution operator then can be written as follows:
\begin{align}
U(t,0)=\sum_{\alpha}e^{i\phi_{\alpha}(t)}|\mu_{\alpha}(t)\rangle\langle\mu_{\alpha}(0)|,
\end{align}
where $\phi_{\alpha}(t)=\int_{0}^{t}dt^{\prime}[A_{\alpha\alpha}(t^{\prime})-H_{\alpha\alpha}(t^{\prime})]$ is the sum of the geometric and dynamical phases.
In NHQC~\cite{Sjoeqvist2012,Xugf2012}, parallel transport conditions $\langle\mu_{\alpha}(t)|H(t)|\mu_{\alpha}(t)\rangle=0$ were applied to the Hamiltonian to remove the dynamical phases, 
while in NHQC+~\cite{Liu2019} scenario, the parallel transport condition was replaced with a more wilder condition: $\int_{0}^{\tau}dt\langle\mu_{\alpha}(t)|H(t)|\mu_{\alpha}(t)\rangle=0$ at the end of the cyclic evolution, 
making it possible for NHQC+ to be compatible with most of the optimization schemes.

\subsection{\label{offset}Offsetting the dynamical phase using a part of the geometric phase}
In this subsection, another way to implement GQC is introduced, where the dynamical phase is offset by a part of the geometric phases\cite{Liu2020a}. 

Consider a set of ancillary states $\{|\mu_{\alpha}(t)\rangle\}$ dependent on two time-varying real parameters: $\lambda_{1}(t)$ and $\lambda_{2}(t)$. 
Then, $|\mu_{\alpha}(t)\rangle=R(\bm{\lambda})|\mu_{\alpha}(0)\rangle$ and $A_{\alpha\beta}(t)=A_{\alpha\beta}^{(\lambda_{1})}(t)+A_{\alpha\beta}^{(\lambda_{2})}(t)$,
where $\bm{\lambda}(t)=\big(\lambda_{1}(t),\lambda_{2}(t)\big)^{\text{T}}$ and 
\begin{align}
A_{\alpha\beta}^{(\lambda_{j})}(t)=\dot{\lambda}_{j}(t)\langle\mu_{\alpha}(\bm{\lambda})|i\frac{\partial}{\partial \lambda_{j}}|\mu_{\beta}(\bm{\lambda})\rangle,
\label{a-lambda-1}
\end{align}
that is, the gauge potential can be divided into two parts: $A^{(\lambda_{1})}(t)$ and $A^{(\lambda_{2})}(t)$. 
If $A^{(\lambda_{1})}(t)$ is equal to the dynamical part of the effective Hamiltonian:
$A_{\alpha\beta}^{(\lambda_{1})}(t)=H_{\alpha\beta}(t)$, 
then the time-evolution operator is governed by a part of gauge potential
$A_{\alpha\beta}^{(\lambda_{2})}(t)=A_{\alpha\beta}(t)-H_{\alpha\beta}(t)$.
For example, consider $R(\bm{\lambda})=e^{X(\bm{\lambda})}$, 
then  
\begin{align}
   A_{\alpha\beta}^{(\lambda_{1})}(t)=\langle\mu_{\alpha}(\bm{\lambda})|\big[i\dot{\lambda}_{1}(t)\frac{\partial X(\bm{\lambda})}{\partial\lambda_{1}}\big]|\mu_{\beta}(\bm{\lambda})\rangle. 
   \label{a-lambda}
\end{align}
By substituting $H(t)=i\dot{\lambda}_{1}(t)\frac{\partial X(\bm{\lambda})}{\partial\lambda_{1}}$ in the effective Hamiltonian $H_{\text{eff}}(t)$, $H_{\alpha\beta}(t)$ will be cancelled out with $A_{\alpha\beta}^{(\lambda_{1})}(t)$. 
The operator $A^{(\lambda_{2})}(t)$ is transformed as a gauge potential under the change 
$|\mu_{\alpha}(t)\rangle\rightarrow\sum_{\beta}|\mu_{\alpha}(t)\rangle V_{\alpha\beta}^{\prime}(\lambda_{2})$, 
where $V^{\prime}(\lambda_{2}(t))$ is any once differentiable operator satisfied $V^{\prime}(\lambda_{2}(\tau))=V^{\prime}(\lambda_{2}(0))$.

For a more specific example, consider $X(t)$ in the following form
\begin{align}
    X\big(\bm{\lambda}(t)\big)=-iF\big(\lambda_{1}(t)\big)H_{0}\big(\lambda_{2}(t)\big),
    \label{sr}
\end{align}
where $F\big(\lambda_{1}\big)$ is a $\lambda_{1}$-dependent real function and $H_{0}(\lambda_{2})$ is a $\lambda_{2}$-dependent hermitian operator. 
To cancel out the dynamical part of the effective Hamiltonian, 
$H(t)= i\dot{\lambda}_{1}(t)\frac{\partial X}{\partial\lambda_{1}}=\dot{\lambda}_{1}(t)\frac{\partial F(\lambda_{1})}{\partial \lambda_{1}}H_{0}(\lambda_{2})$ was selected, 
resulting in a pure geometric effective Hamiltonian 
$H_{\text{eff}}(t)=A^{(\lambda_{2})}(t)=\dot{\lambda}_{2}(t)R^{\dagger}(\bm{\lambda})i\partial_{\lambda_{2}}R(\bm{\lambda})$. Then, Eq. \eqref{derivative}  can be simplified to
\begin{align}
 \frac{d}{dt}{c}_{\alpha \beta}(t)=i\sum_{\gamma}A_{\alpha\gamma}^{(\lambda_{2})}(t)c_{\gamma \beta}(t).
 \label{derivative-2}
\end{align}
The formal solution of Eq. \eqref{derivative-2} can be expressed as follows:
$c(t)=\mathcal{T}\exp[i\int_{0}^{t}A^{(\lambda_{2})}(t^{\prime})dt^{\prime}]$. 
Using the cyclic condition $|\mu_{\alpha}(\tau)\rangle=|\mu_{\alpha}(0)\rangle$ and the definition of $c_{\alpha\beta}(t)$,  
$|\psi_{\alpha}(\tau)\rangle=\sum_{\beta}c_{\alpha\beta}(\tau)|\mu_{\beta}(0)\rangle$,
indicating that $c(\tau)$ is just the transformation matrix from the initial states to the final states. 
Therefore, the temporal evolution operator can be expressed as follows:
\begin{align}
    U(\tau,0)=c(\tau)=\mathcal{T}e^{i\int_{0}^{t}A^{(\lambda_{2})}(t^{\prime})dt^{\prime}}.
    \label{u1}
\end{align}
This cyclic unitary time-evolution operator is geometric because $A^{(\lambda_{2})}(t)$ is a gauge potential. 
\subsection{\label{floquet_gqc}Non-Abelian geometric phases in periodically driven systems}
In this subsection, a brief introduction of the work reported in Ref. \cite{PhysRevA.100.012127} is provided, demonstrating that a non-Abelian geometric phase will form in the adiabatic evolution of a quantum system within a fully degenerate Floquet band.

Consider a driven quantum system with the following Hamiltonian:
\begin{align}
  H(t)=f(\lambda_{1}(t))H_{0}(\lambda_{2}(t)),   
  \label{f-ham}
\end{align}
where $f(\lambda_{1}+T)=f(\lambda_{1})$ is a periodic real function with period $T$ and $\int_{0}^{T}f(\lambda_{1})d\lambda_{1}=0$. 
To remove the dynamical phase using the strategy introduced in Sec. \ref{offset}, $R(\bm{\lambda})=e^{X(\bm{\lambda})}$ should be selected, where $X(\bm{\lambda})$ can be expressed as \eqref{sr}.
$F(\lambda_{1})$ is the primitive function of $f(\lambda_{1})$ and the real function $f(\lambda_{1})$ satisfies the condition: $\int_{0}^{T}f(\lambda_{1})d\lambda_{1}=0$.
It can be verified that $F(\lambda_{1})$ is also periodic with period $T$, so are both $R(\bm{\lambda})$ and $A^{(\lambda_{2})}(\lambda_{1},t)$. 
$A^{(\lambda_{2})}(\lambda_{1},t)$ and $c(\lambda_{1},t)$ can be expanded in a complete set of basis of $\lambda_{1}$ parameter space: $\{e^{i l \lambda_{1}(2\pi/T)}|l\in Z\}$, with which Eq. \eqref{derivative-2} can be recast into:
\begin{align}
    \frac{d}{dt}c_{\alpha\beta}^{(k)}(t)=\sum_{\gamma;m}iH^{F}_{\alpha\gamma;km}(t)c_{\gamma\beta}^{(m)}(t),
    \label{f-derivative}
\end{align}
where $H^{F}(t)\equiv A^{(\lambda_{2})}(\lambda_{1},t)-i\partial_{\lambda_{1}}$, 
$O_{\alpha\gamma;km}(t)=(1/T)\int_{0}^{T}d\lambda_{1}e^{-i(k-m)(2\pi/T)\lambda_{1}}\langle\mu_{\alpha}(\bm{\lambda})|O(t)|\mu_{\gamma}(\bm{\lambda})\rangle$ is the matrix element of operator $O(t)$.
When $\lambda_{1}(t)=\omega t$ and $T=2\pi$, we have 
\begin{align}
    H^{F}_{\alpha\gamma;km}(t)=A_{\alpha\gamma;km}^{(\lambda_{2})}(t)-k\omega\delta_{km}\delta_{\alpha\gamma}.
\end{align}
Clearly, $|\tilde{l}\rangle=e^{il(2\pi/T) \lambda_{1}}$ are the eigenbasis of operator $i\partial_{\lambda_{1}}$ with eigenvalues $-l(2\pi/T)$. 
These eigenvalues do not change when the time $t$ is varied; hence, they form bands, which are referred to as Floquet bands \cite{PhysRevA.100.012127}. 
In other words, the operator $i\partial_{\lambda_{1}}$ is diagonal in basis $|\tilde{l}\rangle$. 
The operator $A^{(\lambda_{2})}(\lambda_{1},t)$ in general is not diagonal in this basis; its off-diagonal terms describe the coupling between different Floquet bands.
However, according to Eq. \eqref{a-lambda-1}, $A^{(\lambda_{2})}_{\alpha\gamma;km}(t)$ is proportional to $\dot{\lambda}_{2}(t)$.
If $\lambda_{2}(t)$ changes slowly in time and $\lambda_{1}(t)$ is a fast time-varying parameter so that $A^{(\lambda_{2})}_{\alpha\gamma;km}(t)\ll k\omega,\ \forall k$,
we may ignore all the terms of $H^{F}(t)$ with $k\neq m$, i.e.,
\begin{align}
  H^{F}_{\alpha,\gamma;k,m}(t)\approx [A_{\alpha\gamma;km}^{(\lambda_{2})}(t)-k\omega\delta_{\alpha\gamma}]\delta_{km}.
  \label{adiabatic_approx}
\end{align}
In analogy to the conventional adiabatic approximation, this is the adiabatic approximation of Floquet band. 
Using Eq. \eqref{adiabatic_approx}, the approximate solution of Eq. \eqref{f-derivative} can be obtained as follows:
\begin{align}
 U(\tau,0)=c(\tau)\approx\mathcal{T}e^{i\int_{0}^{\tau}A^{(\lambda_{2})}_{0}(t)dt},
 \label{u2}
\end{align}
where $A^{(\lambda_{2})}_{0}(t)=(1/T)\int_{0}^{T}A^{(\lambda_{2})}(\lambda_{1},t)d\lambda_{1}$. 

By comparing Eq. \eqref{u2} with Eq. \eqref{u1}, it can be clearly observed that $A^{(\lambda_{2})}(\lambda_{1},t)$ can be replaced with an average quantity $A^{(\lambda_{2})}_{0}(t)$, which can be interpreted as a gauge potential resulting from the adiabatic motion in a single degenerate Floquet band. 
However, it is still not clear what benefits this replacement will provide to quantum computing tasks and whether a universal quantum computation scheme can be constructed in practice using the gauge potential $A^{(\lambda_{2})}_{0}(t)$. 
We will address these problems in the next section. 

\begin{figure}
    \centering
    \includegraphics[width=0.43\textwidth]{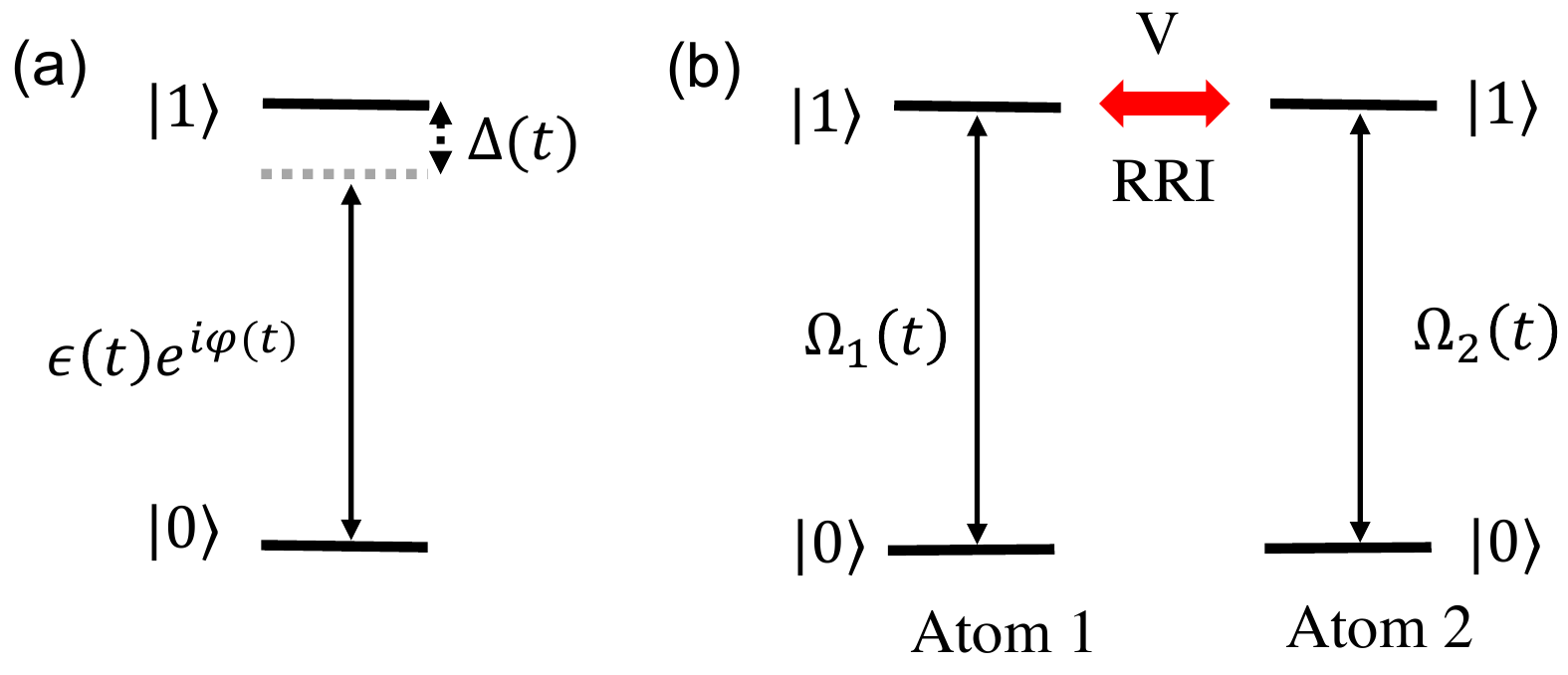}
    \caption{ (a) Setup for two-level Rydberg atom. The two-level system is driven off-resonantly with time-dependent detuning $\Delta(t)$, Rabi frequency $\epsilon(t)$, and phase $\varphi(t)$. 
    (b) Illustration of control of two-qubit gate based on RRI between two identical two-level Rydberg atoms, where $V$ is the RRI strength. For individual Rydberg atoms, a resonant microwave pulse with Rabi frequency $\Omega_{\alpha}$ can be applied to facilitate the transition $|0\rangle\rightarrow|1\rangle$.}
    \label{fig:control-illust}
\end{figure}

\begin{figure}
    \centering
    \includegraphics[width=0.48\textwidth]{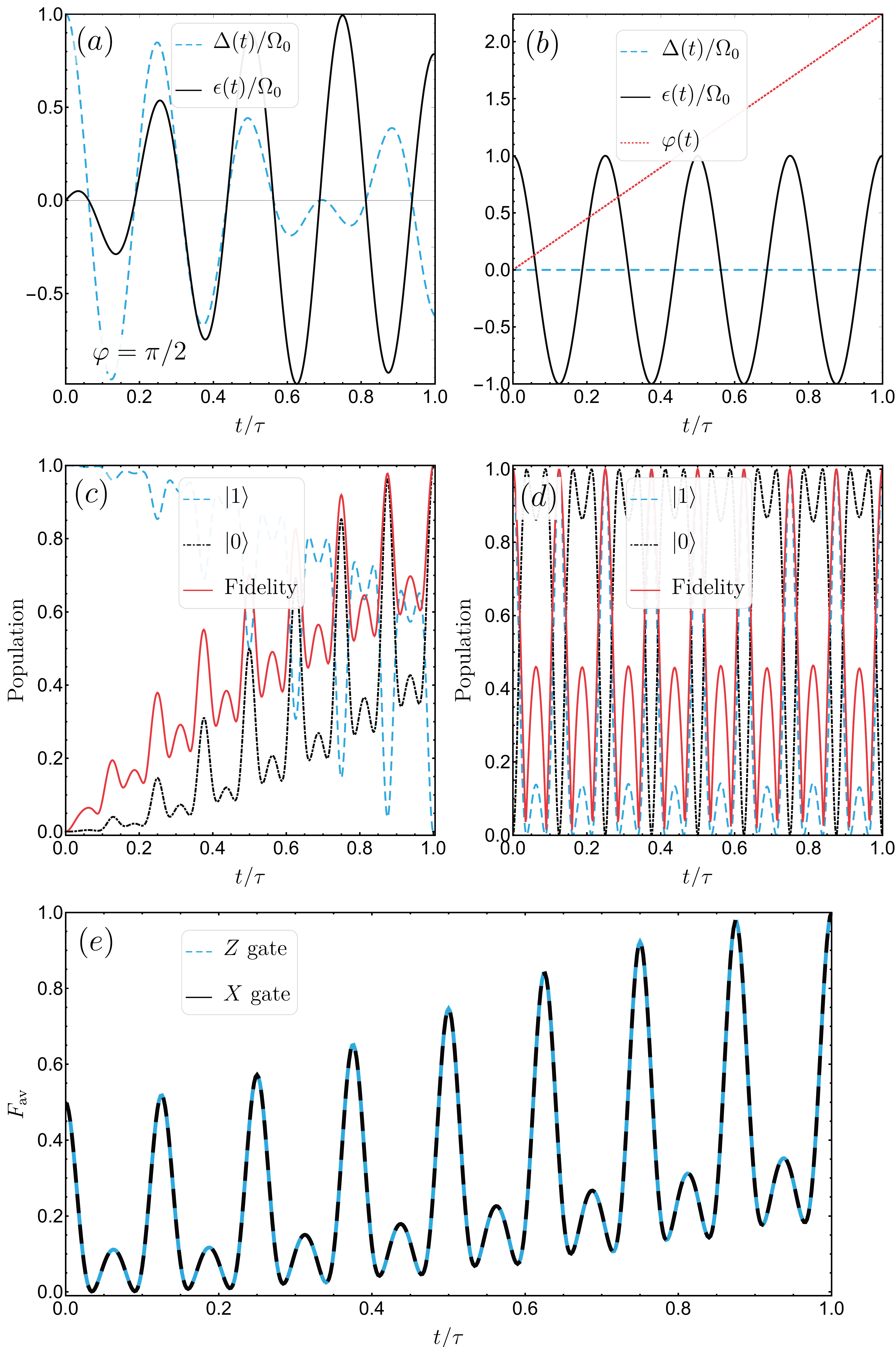}
    \caption{(a) Detuning $\Delta(t)$ (blue dashed line) and Rabi frequency $\epsilon(t)$ (black solid line) for the numerical simulation of the X gate using FGQC. 
    (b) Detuning $\Delta(t)$ (blue dashed line), Rabi frequency $\epsilon(t)$ (black solid line), and phase $\varphi(t)$ (red dotted line) for the numerical simulation of the Z gate using FGQC.
    Temporal evolution of populations (blue dashed line for state $|1\rangle$, black dot dashed line for state $|0\rangle$) and fidelities (red solid line)  with a given initial state $|\psi(0)\rangle=|1\rangle$ for the FGQC X (c) and Z gates (d). 
    (e) Temporal evolution of gate fidelities for the X (black solid line) and Z gates (blue dashed line).}
    \label{fig:pulses}
\end{figure}

\begin{figure}
    \centering
    \includegraphics[width=0.45\textwidth]{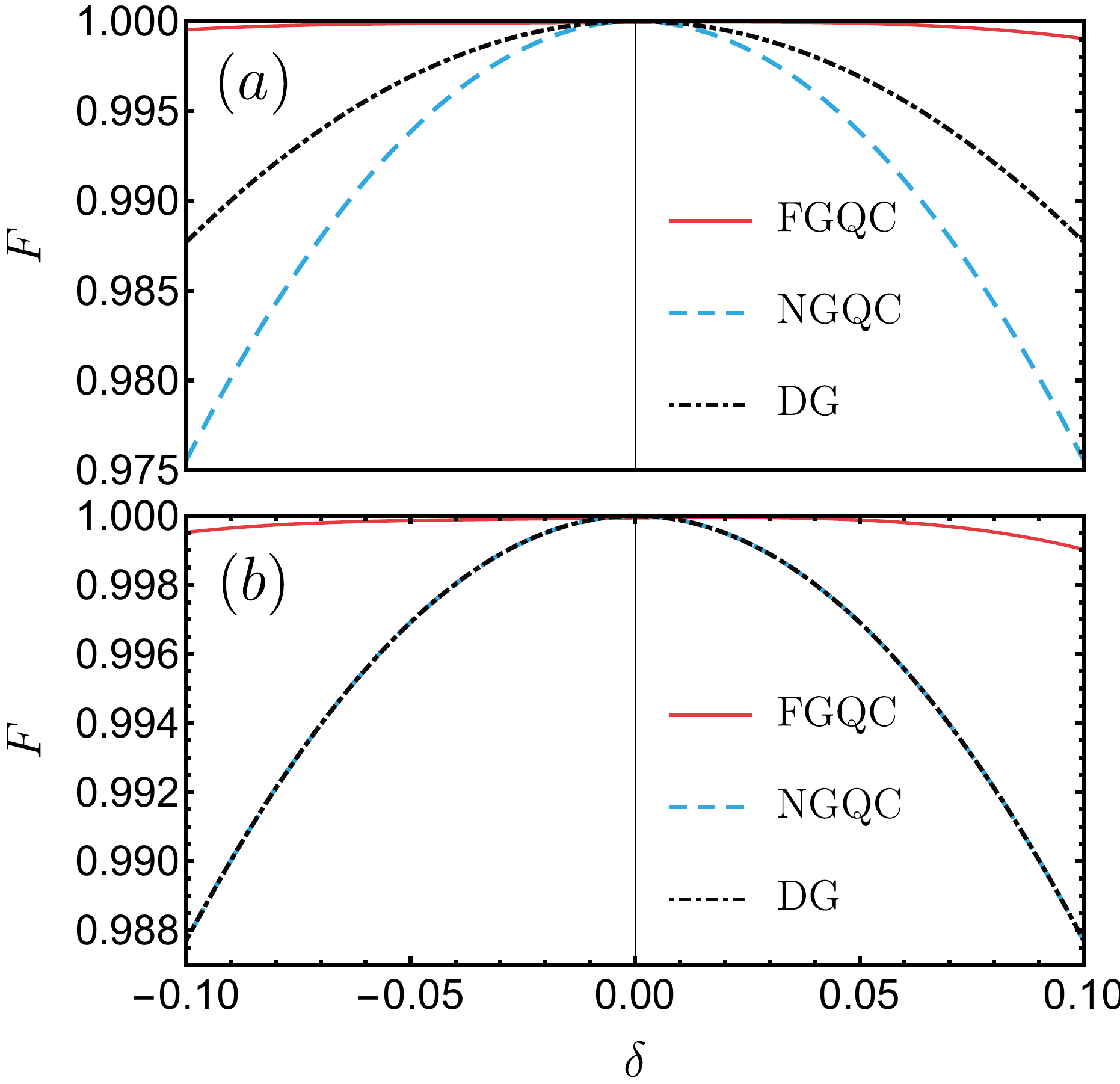}
    \caption{Fidelities of the Z (a) and X (b) gates vs. the amplitude of the global control error (without considering decoherence in the simulation). 
    The results of three different protocols are shown:  FGQC (red solid line), NGQC (blue dashed line), and standard DG (black dot dashed line).}
    \label{fidelity}
\end{figure}

\section{\label{sec:physics}Floquet GQC based on Rydberg atoms}
For universal quantum computation (QC), at least two types of noncommutable single-qubit gates and one nontrivial two-qubit gate are needed \cite{Lloyd1995}. 
On the other hand, the Rydberg atom provides an appealing experimental platform for the implementation of quantum computation because of its long coherence time \cite{Saffman2010,Saffman2016}. 
In this section, using the gauge potential $A^{(\lambda_{2})}_{0}(t)$ in Eq. \eqref{u2},  we give a possible physical realizations of universal single-qubit gates and a nontrivial two-qubit gate based on Rydberg atoms. They constitute a universal geometric QC scheme.
This geometric QC scheme is referred to as Floquet geometric quantum computation (FGQC). Notably, our theory can be applied to other platforms; only Rydberg atom has been used to show the feasibility.

\subsection{Single-qubit gates}

Consider a Rydberg atom with the ground state and the Rydberg state encoding $|0\rangle$ and $|1\rangle$ respectively.
By applying appropriately selected laser light,
the ground state can be directly coupled to the Rydberg state directly using one photon process or via multi-photon process using one or two intermediate (non-Rydberg) states \cite{omran570}. 
After performing an adiabatic elimination of the intermediate state, 
the Hamiltonian in rotating frame can be written as follows:
\begin{align}
   H(t)=\frac{\Delta(t)}{2}\sigma_{z}+\frac{\epsilon(t)}{2}[\cos\varphi(t)\sigma_{x}+\sin\varphi(t)\sigma_{y}].
   \label{eq:control}
\end{align}
This Hamiltonian drives the transition $|0\rangle\rightarrow |1\rangle$ with Rabi frequency $\epsilon(t)\exp[i\varphi(t)]$, as shown in Fig. \ref{fig:control-illust} (a), where $\Delta(t)$ is the effective detuning.
In an actual experiment, the laser parameters can be modulated with an acousto-optic modulator (AOM) driven by an arbitrary waveform generator (AWG). By combining with some feed-forward approaches, the experimentally applied pulse can be ensured to be a faithful representation of the desired profile\cite{Saffman2010,Saffman2016,omran570}.

By taking ${|0\rangle, |1\rangle}$ as the computational basis, an arbitrary one-qubit nonadiabatic geometric gate can be achieved as follows:
\begin{align}
    U(\tau,0)=e^{i\gamma\mathbf{F}\cdot\mathbf{n}},
    \label{u_aim}
\end{align}
where $\mathbf{F}=\frac{1}{2}(\sigma_{x},\sigma_{y},\sigma_{z})^{\text{T}}$, $\mathbf{n}=(\sin\theta\cos\phi,\sin\theta\sin\phi,\cos\theta)$ is an arbitrary unit vector and $\gamma$ is an arbitrary phase with geometric feature. Eq. \eqref{u_aim} describes a rotational operation around the axis $\mathbf{n}$ by an angle $\gamma$.
If the pulse shapes of the following form are selected:
\begin{align}
   \Delta(t)&=\Delta_{0}(t)\cos(\omega t+\theta_{0}), \\
   \epsilon(t)&=\epsilon_{0}(t)\cos(\omega t+\theta_{0}),
\end{align}
where $\omega$ and $\theta_{0}$ are time-independent parameters; $\Delta_{0}(t)$ and $\epsilon_{0}(t)$ are in general time-dependent real functions. 
Then, the Hamiltonian can be rewritten as follows:
\begin{align}
    H(t)=\cos(\omega t+\theta_{0})\mathbf{F}\cdot\mathbf{r}(t),
    \label{single-ham}
\end{align}
where $\mathbf{r}(t)=(\epsilon_{0}(t)\cos\varphi(t),\epsilon_{0}(t)\sin\varphi(t),\Delta_{0}(t))^{\text{T}}$.  
Eq. \eqref{single-ham} has the same form as Eq. \eqref{f-ham} if $\lambda_{1}(t)=\omega t+\theta_{0}$, $f\big(\lambda_{1}(t)\big)=\cos[\lambda_{1}(t)]$ and $H_{0}(t)=\mathbf{F}\cdot\mathbf{r}(t)$.
According to Eq. \eqref{sr}, for a FGQC scenario, the ancillary basis is given by the transformation: $R(\omega t,t)=\exp\big[-i\frac{\sin(\omega t+\theta_{0})}{\omega}\mathbf{F}\cdot\mathbf{r}(t)\big]$. 
This transformation will result in the following approximated effective Hamiltonian:
\begin{align}
    A^{(\lambda_{2})}_{0}(t)
    =\Omega(t)[1-J_{0}(|\mathbf{r}(t)|/\omega)]\mathbf{F}\cdot\mathbf{n}(t),
    \label{eff_hamiltonian}
\end{align}
where $\Omega(t)\mathbf{n}(t)\equiv\mathbf{r}(t)\times\dot{\mathbf{r}}(t)/|\mathbf{r}(t)|^{2}$ with $\mathbf{n}(t)$, a unit vector; $J_{0}(a)$ is the zero-order Bessel function. 
If the plane determined by vectors $\mathbf{r}$ and $\dot{\mathbf{r}}$ is fixed, 
then $\mathbf{n}(t)$ does not change, and the effective Hamiltonian \eqref{eff_hamiltonian} at different times can commutable with each other. 
The time evolution operator \eqref{u2} can be simplified into $U(\tau,0)=\exp\big[i\gamma(\tau)\mathbf{F}\cdot\mathbf{n}(t)\big]$ with $\gamma(\tau)=\int_{0}^{\tau}dt \Omega(t)\big[1-J_{0}(|\mathbf{r}(t)|/\omega)\big]$. $U(\tau,0)$ has the same form as \eqref{u_aim}. As $\mathbf{n}$ and $\gamma(\tau)$ can take any value, a universal single-qubit gate can be applied using FGQC scenario. 

Taking the Z and X gates as examples, we will give a possible set of pulse shapes and parameters for each gate. 
Moreover, to evaluate the performance of these gates with the given parameters, numerical results in the presence of both decoherence and a certain kind of systematic error will be presented (we also presented similar contents for the Hadamard and T gates in Appendix \ref{sec:T_H}).   
To implement a single-qubit Z gate using FGQC, we selected $\Delta_{0}(t)=0$,  $\epsilon_{0}(t)=\Omega_{0}$ and $\varphi(t)=Nt$, with $\Omega_{0}$ and $N$ are real constants which satisfy $N[1-J_{0}(|\mathbf{r}(t)|/\omega)]\tau=\pi$, and $\tau$ is the run time of the gate. Here, $|\mathbf{r}(t)|=\Omega_{0}$; the pulse shapes are shown in Fig. \ref{fig:pulses} (a).
In the numerical simulation, we set $\Omega_{0}=2\times 2\pi$ MHz, $\omega\approx 0.513\times 2\pi$ MHz, $N\approx 45.728\times2\pi$ KHz, and the run time $\tau\approx 7.797$ $\mu$s. 
For the X gate, $\epsilon_{0}(t)=\Omega_{0}\sin Mt$, $\Delta_{0}(t)=\Omega_{0}\cos Mt$, and $\varphi(t)=\pi/2$ with $M=N$ real time-independent constants.  
The corresponding pulse shapes are shown in Fig. \ref{fig:pulses} (b). 
Other parameters are the same as the Z gate.

The noisy gate operation in the presence of global control error is described by the master equation as follows:
\begin{align}
    \frac{d\rho(t)}{dt}=-i[H^{\prime}(t),\rho(t)]+\sum_{i=1}^{2}\mathcal{L}_{i}[\rho(t)],
    \label{eq:m-eq}
\end{align}
where $H^{\prime}(t)=(1+\delta)H(t)$ with $H(t)$ the ideal Hamiltonian, $\delta$ represents the amplitude of global control error, $\mathcal{L}_{i}[\rho]=\gamma_{i}[A_{i}\rho A_{i}^{+}-(1/2)\{A_{i}^{+}A_{i},\rho\}]$ is the Lindblad superoperator which acts on the density matrix $\rho$ of the quantum system; $\{a,b\}=ab+ba$ is the anticommutator.
In the case of a two-level Rydberg atom, the incoherent processes, including decay and dephasing: $A_{1}=\sigma_{-}$, $A_{2}=\sigma_{z}$, where $\sigma_{-}$ is the spin ladder operator and $\sigma_{z}$ is the Pauli $z$ matrix; 
thus, $\gamma_{1}$ and $\gamma_{2}$ are the decay and dephasing rates, respectively. 
Here, we selected $\gamma_{1}=12.83$ Hz, $\gamma_{2}=128.3$ Hz and $\delta=0.05$.

In Figs. \ref{fig:pulses} (c) and (d), for the X and Z gates, the temporal evolution of state populations is shown for $|0\rangle$ and $|1\rangle$ with a given initial state $|\psi(0)\rangle=|1\rangle$.
The corresponding fidelities between the target states and the temporal evolution states are also shown in these figures; the final state fidelities for the X and Z gates at $\tau$ are as follows: $F_{X}(\tau)\approx 0.9997$ and $F_{Z}(\tau)\approx 0.9998$ respectively.  
The gate fidelity was also investigated, which is defined by $F=(1/2\pi)\int_{0}^{2\pi}\langle\psi_{I}|\rho|\psi_{I}\rangle d\Theta$, for the initial states of the form $|\psi\rangle=\cos\Theta |0\rangle+\sin\Theta |1\rangle$, where a total of 101 different values of $\Theta$ were uniformly selected in the range $[0,2\pi]$. The results are shown in Fig. \ref{fig:pulses} (e). 
The gate fidelities at $\tau$ are $F_{X,\text{gate}}\approx F_{Z,\text{gate}}\approx 0.9992$ for the X and Z gates, respectively. 
These fairly high state and gate fidelities indicate two significant factors: firstly, the run time of our proposal is short enough to lighten the destructive effect of decoherence; secondly, other parameters in the proposal match well with the requirements of FGQC theory. 

With respect to the reachability of our proposal, a recent experimental study in Ref. \cite{omran570} was selected for comparison. 
In this study, the amplitude of $\displaystyle \epsilon ( t)$ can be changed from 0 to $\displaystyle 5\times 2\pi $ MHz within 0.5 $\displaystyle \mu $s, and the detuning $\displaystyle \Delta ( t)$ can be changed from $\displaystyle -15\times 2\pi $ MHz to a value larger than $\displaystyle 7.5\times 2\pi $ MHz within 1 $\displaystyle \mu $s. 
Besides, the pulses in this study were given by a numerical optimization method known as RedCRAB; their shapes are not that smooth. 
While in our proposal, taking the X gate as an example, the parameters $\displaystyle \epsilon ( t)$ and $\displaystyle \Delta ( t)$ should be changed smoothly within the range $[-2 , 2]\times 2\pi\ \text{MHz}$ in about 7.797 $\displaystyle \mu $s; clearly, the pulses in this proposal are more smoother and changes slower within a smaller range. 
In other words, by comparing with recently experimental results, this proposal needs a series of smoother, more slowly varying pulses with smaller ranges. 
Therefore, the control in implementing FGQC X gate is achievable
The control of FGQC Z gate is also slow and smooth, despite $\displaystyle \Delta ( t) =0$, it also needs $\varphi(t)=Nt$, that is $\varphi(t)$ increase linearly at a slow rate, to our knowledge, it is not clear weather this kind of control is achievable.

\subsection{\label{2-gate}Two-qubit gate}

It is shown that an arbitrary one-qubit FGQC gate can be obtained by addressing an individual Rydberg atom with laser pulses. 
To achieve universal GQC, a nontrivial two-qubit gate is needed beside one-qubit gates.
We here demonstrate how to achieve a nontrivial two-qubit FGQC gate using the Rydberg-Rydberg interaction (RRI).

Consider two two-level Rydberg atoms with RRI $V$, as shown in Fig. \ref{fig:control-illust} (b). 
The $\alpha$th atom is resonantly driven by a microwave pulse to achieve the transitions $|0\rangle_{\alpha}\rightarrow |1\rangle_{\alpha}$, with Rabi frequency $\Omega_{\alpha}(t)$. 
In the rotating frame, the Hamiltonian of two-Rydberg-atom system can be expressed as follows~\cite{Zhao2017}:
\begin{equation}
H_{12} =H_{1} \otimes \mathbb{I}_{2} +\mathbb{I}_{1} \otimes H_{2} +V|11\rangle \langle 11|,
\end{equation}
where $\displaystyle H_{\alpha} =\Omega _{\alpha }(t)( |0\rangle _{\alpha } \langle 1|+|1\rangle _{\alpha } \langle 0|)$ is a single-atom Hamiltonian describing the interaction between the $\displaystyle \alpha $th atom and laser pulses; 
$\mathbb{I}_{\alpha}$ is the identity operator acting on the $\alpha$th Rydberg atom.

To achieve a nontrivial two-qubit gate, the Rabi frequencies of laser pulses can be taken as $\displaystyle \Omega _{1}( t) =-\Omega _{R}( t)\cos( \phi /2)$ and $\displaystyle \Omega _{2}( t) =\Omega _{R}( t)\sin( \phi /2)$. Thus,
\begin{align}
H_{12}( t) =\Omega _{R}( t)\left( |B\rangle \langle 00|-|B^{\prime } \rangle \langle 11|+\text{H.c.}\right) +V|11\rangle \langle 11|,
\end{align}
where $\displaystyle |B\rangle =\sin( \phi /2) |01\rangle -\cos( \phi /2) |10\rangle $ and $\displaystyle |B^{\prime } \rangle =\cos( \phi /2) |01\rangle -\sin( \phi /2) |10\rangle $.

Further, taking a rotation $\displaystyle U=\exp[ -iVt|11\rangle \langle 11|]$, then the two-atom Hamiltonian can be recast as follows:
\begin{equation}
H_{\text{rot}}( t) =\Omega _{R}( t)\left( |B\rangle \langle 00|-|B^{\prime } \rangle \langle 11|e^{-iVt}\right) +\text{H.c.}.
\label{eq:h-b}
\end{equation}
If $\displaystyle V\gg \Omega _{R}( t)$, the off-resonant terms are negligible, the simultaneous excitation of two atoms from ground state to  Rydberg states is inhibited; therefore,
\begin{equation}
\ H_{\text{rot}}( t) \approx \Omega _{R}( t) |B\rangle \langle 00|+\text{H.c.}.
\label{h_rot}
\end{equation}
According to Eq. \eqref{h_rot}, the effective Hamiltonian only affects subspace $S=\text{Span}\{|B\rangle ,|00\rangle \}$. 
Therefore, $S$ was set as the computational space. 
Consider $\Omega _{R}( t) =[\Omega _{0}/2]f(\omega t)\exp[i\varphi(t)] $ where $\Omega _{0}$ is a real constant. 
Both $f(\omega t)$ and $\varphi(t)$ are real functions. 
The Hamiltonian \eqref{h_rot} can be rewritten as follows:
\begin{align}
H_{\text{rot}}(\omega t ,t) =f(\omega t)\mathbf{F} \cdot \mathbf{r}^{\prime}( t) ,
\end{align}
where $\displaystyle \mathbf{F} \equiv ( \tilde{\sigma} _{x} ,\tilde{\sigma} _{y} ,\tilde{\sigma} _{z})^{T} /2$ and $\mathbf{r}^{\prime}( t) =\Omega _{0}(\cos \varphi ( t) ,-\sin \varphi ( t) ,0)^{T}$ with $\tilde{\sigma}_{x} \equiv |B\rangle\langle 00|+|00\rangle\langle B|$, 
$\tilde{\sigma}_{y} \equiv -i|B\rangle\langle 00|+i|00\rangle\langle B|$ and 
$\tilde{\sigma}_{z} \equiv |B\rangle\langle B|-|00\rangle\langle 00|$.
To be more specific, we considered $f(\omega t) =\cos \omega t$; then, $\displaystyle H_{\text{rot}}( \omega t,t) =\cos \omega t\mathbf{F} \cdot \mathbf{r}^{\prime}( t)$. 
To construct a FGQC gate, select $R( t) =\exp[ -i\Omega_{0}\sin \omega t\mathbf{F} \cdot \mathbf{r}^{\prime}( t) /\omega ]$, 
satisfying the boundary condition $\displaystyle R( \tau ) =R( 0) =I$ if $\displaystyle \omega \tau =k\pi, \forall k\in N^{+} $. 
The corresponding effective Hamiltonian can be written as follows:
\begin{align}
H_{\text{eff}}(t) =\dot{\varphi }( t) R^{\dagger }( t)( -i\partial /\partial \varphi ) R( t).
\end{align}
Similar to the derivation of Eq. \eqref{u2}, if $\omega\gg \dot\varphi(t)$, then at $\displaystyle t=\tau $, the time evolution operator can be approximated as follows:
\begin{equation}
U( \tau ,0) \approx \mathcal{T} e^{-i\int _{0}^{\tau } H^{( 0)}_{\text{eff}}( t) dt},
\label{eq:two_u}
\end{equation}
where $H^{( 0)}_{\text{eff}}( t)=[ 1-J_{0}( \Omega_{0}/\omega )]\mathbf{F} \cdot (\mathbf{r}^{\prime} \times \dot{\mathbf{r}}^{\prime})/{\Omega_{0}^{2}}$.
If $\displaystyle \varphi ( t) =Nt$, then, $\displaystyle \mathbf{r}^{\prime} \times \dot{\mathbf{r}}^{\prime} /\Omega_{0}^{2} =-N( 0,0,1)^{T}$ and $\displaystyle H^{( 0)}_{\text{eff}}( t) =-N[ 1-J_{0}( \Omega_{0}/\omega )] \tilde{\sigma} _{z} /2$. 

In terms of the basis $\displaystyle \{|00\rangle ,|01\rangle ,|10\rangle ,|11\rangle \}$, the matrix form of average effective Hamiltonian can be expressed as follows: 
\begin{equation}
H^{( 0)}_{\text{eff}}( t) =C\cdot\begin{pmatrix}
0 & 0 & 0 & 0\\
0 & \cos^{2}\left(\frac{\phi }{2}\right) & \frac{-\sin \phi }{2} & 0\\
0 & \frac{-\sin \phi }{2} & \sin^{2}\left(\frac{\phi }{2}\right) & 0\\
0 & 0 & 0 & -1
\end{pmatrix},
\end{equation}	  
where $\displaystyle C \equiv -( N/2)[ 1-J_{0}( \Omega_{0}/\omega )]$. 
The time evolution operator is given by
\begin{align}
U( \tau ,0) & =\begin{pmatrix}
1 & 0 & 0 & 0\\
0 & \frac{1}{2}\left( D_{1} -D_{2}\cos\phi \right) & \frac{1}{2}D_{2}\sin \phi  & 0\\
0 & \frac{1}{2}D_{2}\sin \phi  & \frac{1}{2}\left( D_{1} +D_{2}\cos\phi\right) & 0\\
0 & 0 & 0 & e^{iC \tau }
\end{pmatrix},
\end{align}	 
where $D_{1}=1+\exp(-iC\tau)$ and $D_{2}=1-\exp(-iC\tau)$.
Clearly, when $\displaystyle e^{iC \tau } =-1$ and $\displaystyle \phi =\pi /2$, we have
\begin{equation}
U( \tau ,0) =\begin{pmatrix}
1 & 0 & 0 & 0\\
0 & 0 & 1 & 0\\
0 & 1 & 0 & 0\\
0 & 0 & 0 & -1
\end{pmatrix}. 
\label{eq:swap}
\end{equation}
This operation is a SWAP-like gate because: 
\begin{align}
U( \tau ,0) |10\rangle  & =|01\rangle, \\
U( \tau ,0) |01\rangle  & =|10\rangle ,
\end{align}
this is a nontrivial two-qubit gate ensuring that universal quantum computation can be applied using FGQC.

\begin{figure}
    \centering
    \includegraphics[width=0.43\textwidth]{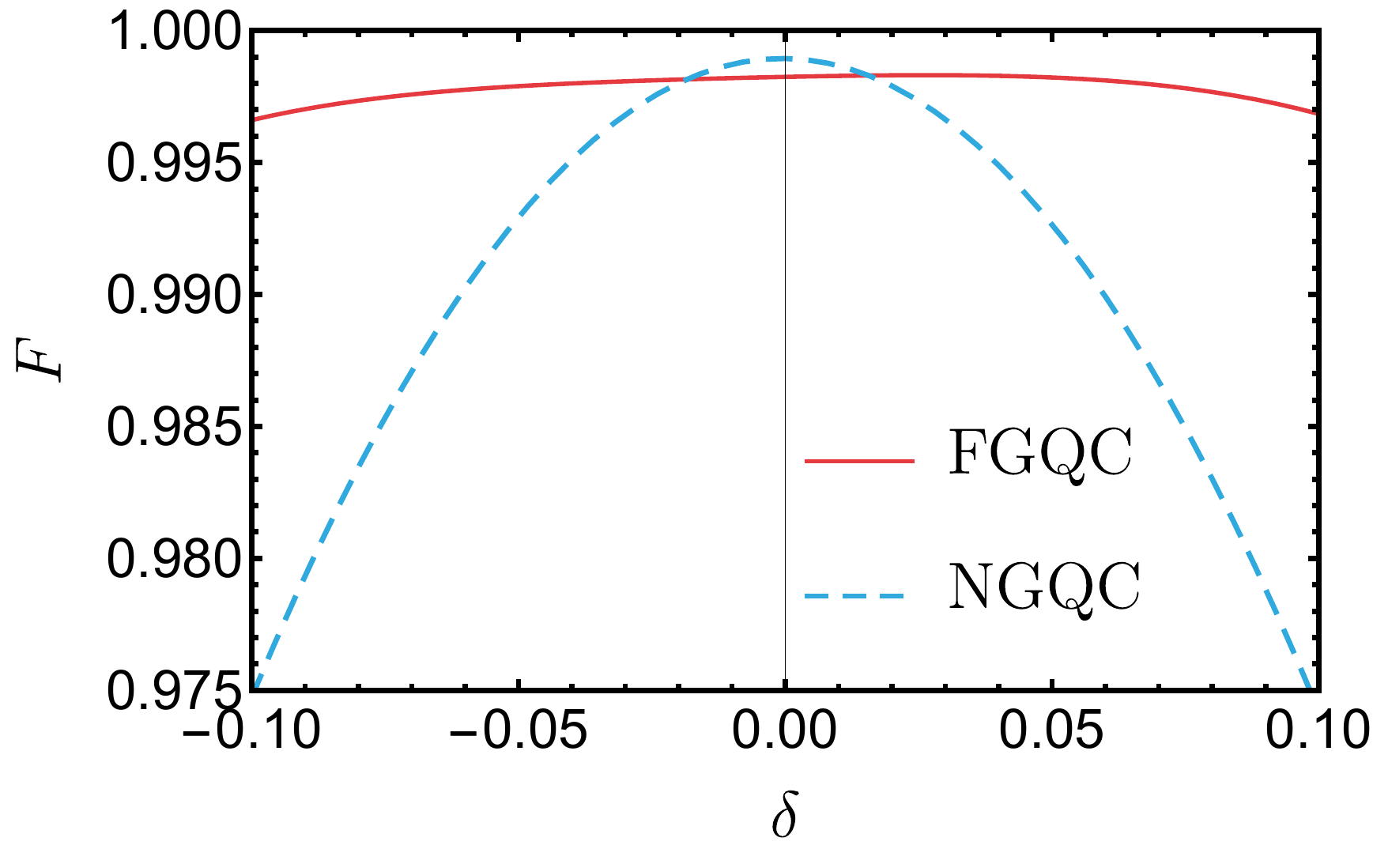}
    \caption{The fidelity $F$ vs. the amplitude $\delta$ of global control error for the same two-qubit gate using FGQC (red solid line) and NGQC (blue dashed line), respectively. The two-qubit gate is given by Eq. \eqref{eq:swap}.}
    \label{fig:two-qubit}
\end{figure}

\subsection{\label{sec:robustness}The robustness of FGQC against global control error}
  
Since only a gauge potential $A^{(\lambda_{2})}_{0}(t)$ can be attributed to the cyclic unitary evolution operator \eqref{u2}, FGQC is a geometric scenario. 
However, as shown in Ref. \cite{Liu2020}, a geometric scenario is not necessarily resistant to global control error. 
Here, we will demonstrate that the FGQC is a resistant GQC scenario to global control error.
To prove this, we consider a quantum system which is subject to global control error. 
Given the initial state $|\psi_{m}(0)\rangle$, 
the corresponding time evolution state can be expressed as $|\psi_{m}^{\prime}(\tau)\rangle=U^{\prime}(\tau,0)|\psi_{m}(0)\rangle$ with 
$U^{\prime}(\tau,0)=\mathcal{T}\exp[i\int_{0}^{\tau}dt(A^{(\lambda_{2})}(\omega t,t)-\delta H(\omega t,t))]$. 
Normally, $\delta$ is a small quantity in experiment and it is reasonable to assume that $\delta H_{\alpha\gamma}(t)\ll k\omega$.
Similar to the derivation of Eq. \eqref{u2}, $U^{\prime}(\tau,0)$ can be approximately expressed as follows:
\begin{align}
    U^{\prime}(\tau,0)\approx\mathcal{T}e^{i\int_{0}^{\tau}[A^{(\lambda_{2})}_{0}(t)-\delta \bar{H}(t)]dt},
    \label{eq:u-error}
\end{align}
where $\bar{H}(t)=(1/T)\int_{0}^{2\pi}H(\lambda_{1},t)d\lambda_{1}$;
then, using perturbation theory up to $\mathcal{O}(\delta)$:
\begin{align}
    |\psi_{m}^{\prime}(\tau)\rangle=|\psi_{m}(\tau)\rangle-\delta\cdot i\sum_{n}|\psi_{n}(\tau)\rangle Q_{nm}(\tau)+\mathcal{O}(\delta^{2}),
    \label{error-state}
\end{align}
where $|\psi_{m}(\tau)\rangle=U(\tau,0)|\psi_{m}(0)\rangle$ is the ideal time evolution state and $Q_{nm}(t)=\int_{0}^{t}dt^{\prime}\langle\psi_{n}(t^{\prime})|\bar{H}(t^{\prime})|\psi_{m}(t^{\prime})\rangle$. 

Eq. \eqref{error-state} clearly shows that to eliminate the second term at the right side, $Q_{mn}(\tau)$ should be equal to zero:
\begin{align}
    \int_{0}^{\tau}\langle\psi_{n}(t)|\bar{H}(t)|\psi_{m}(t)\rangle dt=0\ \ \forall m,n.
    \label{condition}
\end{align}
Eq. \eqref{condition} maintains the robustness of FGQC against global control error to the first order in $\delta$.
In a FGQC scenario, $\bar{H}$ is given by 
\begin{align}
\bar{H}(t)=\bar{f}(\lambda_{1})H_{0}(\lambda_{2}),
\end{align}
where $\bar{f}(\lambda_{1})=(1/T)\int_{0}^{T}f(\lambda_{1})d\lambda_{1}$ and $f(\lambda_{1})$ is a periodic function with period $T$. 
In this study, $f(\lambda_{1})=\cos(\lambda_{1})$, then, $\bar{f}(\lambda_{1})$ is equal to zero, $\bar{H}(t)=0$, and Eq. \eqref{condition} is satisfied. 
The above analysis is based on single-qubit FGQC theory, but it can also be applied to the two-qubit gate case. 
For two-qubit case, one only need replace $A_{0}^{(\lambda_{2})}(t)$ and $H(t)$ with $H_{\text{eff}}^{(0)}(t)$ and $H_{\text{rot}}(t)$ respectively.
Therefore, FGQC is resistant to global control error.

Numerical evidence of the robustness of FGQC against static global control error. 
In Fig. \ref{fidelity} and Fig. \ref{fig:two-qubit}, numerical results of the single-qubit $X$, $Z$ gates and the two-qubit gate described in Eq. \eqref{eq:swap} are given.
They show the single- and two-qubit gate fidelities vs. $\delta$, respectively; 
clearly, the FGQC scenario is more robust than the standard dynamical scenario and the NGQC scenario in solving global control errors (see Appendix \ref{sec:detail} for the detail of the standard dynamical and NGQC gates). 

The parameters of the FGQC two-qubit gate were selected as follows: $\Omega_{0}=2\times 2\pi$ MHz, $\omega\approx 0.52\times 2\pi$ MHz, $N\approx 0.58$ MHz, $V\approx 0.33$ GHz and the run time $\tau\approx 7.7 \mu$s.
Because the presence of the $|B^\prime\rangle\langle 11|$ term and its hermitian conjugation, the fidelity at $\delta=0$ is not 1.

The single-qubit FGQC T and Hadamard gates were also simulated, the results are shown in Appendix \ref{sec:T_H}

When the error become time-dependent. For a time dependent error $\delta=\delta(t)$, we should modify the unitary time evolution operator in  Eq. \eqref{eq:u-error} by replacing $\delta\bar{H}(t)$ in the exponent with the following expression:
\begin{align}
(1/T)\int_{0}^{2\pi}\delta(t)H(\lambda_{1},t)d\lambda_{1}.
\label{eq:noname}
\end{align}
In our proposal, $\lambda_{1}=\omega t$ and $H(\lambda_{1},t)=\cos(\omega t)H_{0}(t)$ with $\omega$ a time-independent real number and $H_{0}(t)$ is a slowly varying operator, 
the expression Eq. \eqref{eq:noname} can be rewritten as $(1/T)\int_{0}^{2\pi} dt\cos(\omega t)\delta(t)H_{0}(t)$. Clearly, when $\delta(t)$ is slow enough, this expression is approximately equal to zero. 
Therefore, FGQC is still a robust scenario when $\delta(t)$ is slow enough.

\begin{figure}
    \centering
    \includegraphics[width=0.43\textwidth]{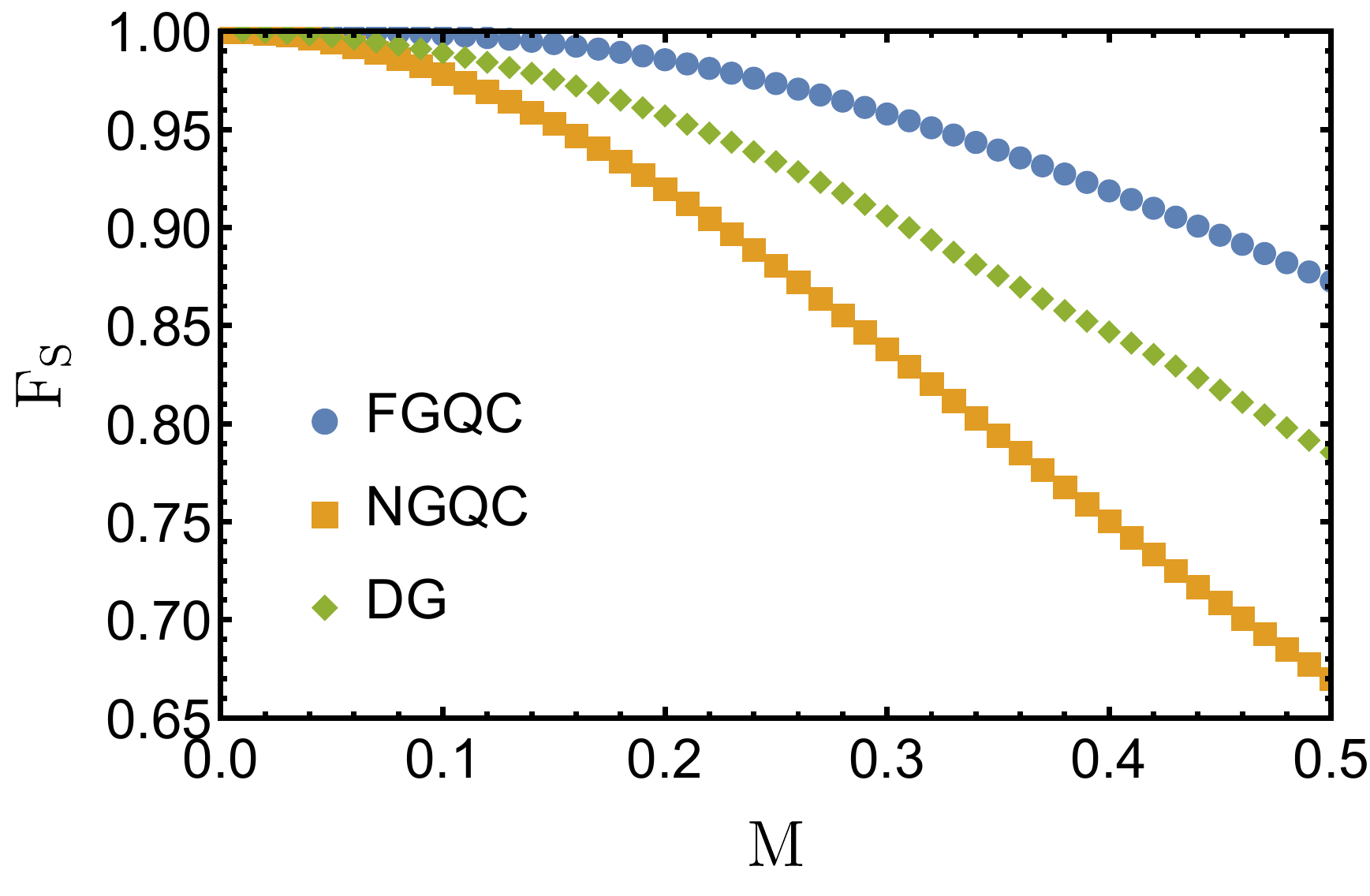}
    \caption{The average gate fidelity $F_{S}$ vs. $M$ for the Z gate based on FGQC (blue dots), NGQC (yellow rectangle) and DG (green diamond), respectively.}
    \label{fig:fluc_fid}
\end{figure}

A numerical demonstration was also given through simulating a Z gate in the presence of $\delta(t)$.  
A stochastic process called Ornstein–Uhlenbeck (O-U) process was considered to describe the time-dependence of $\delta(t)$. 
To be specific, O-U process is a stochastic process with power spectral density (PSD) $S(\omega)=\frac{\Gamma}{\pi(\Gamma^{2}+\omega^{2})}$, where $1/\Gamma$ is the correlation time. 
The Hamiltonian is given by:
\begin{align}
    H^{\prime}(t)=[1+M\delta(t)]H(t),
\end{align}
where $H(t)$ is the ideal Hamiltonian, $M$ is a zoom factor.
In the simulation, we set $\Gamma \tau=0.001$. For each $M$; we firstly sampled 100 such process randomly: $\{\delta_{i}(t)|i=1,\cdots,100\}$; then, we calculate the gate fidelity for each $\delta_{i}(t)$ and obtain the average gate fidelity for these $100$ processes. The results are shown in Fig. \ref{fig:fluc_fid}.

\section{\label{sec:conclusion}Conclusion}
A new geometric scheme called FGQC is reported. In a FGQC scenario, error-resilient geometric gates based on periodically driven two-level systems can be constructed via a new non-Abelian geometric phase. 
This non-Abelian geometric phase emerges from a periodically driven quantum systems and was found in a recent study~\cite{PhysRevA.100.012127}. 
To construct FGQC, possible implementations of universal single-qubit gates and a nontrivial two-qubit gate using Rydberg atoms were proposed. 
To numerically evaluate its realistic performance, the X and Z gates were simulated in the presence of decoherence and global control error using recent experimental parameters. 
The noisy X and Z gates acting on a given initial state $|\psi(0)\rangle=|1\rangle$ reached state fidelities as high as: $F_{X}(\tau)\approx 0.9997$ and $F_{Z}(\tau)\approx 0.9998$; the gate fidelities for the X and Z gates are $F_{X,\text{gate}}\approx F_{Z,\text{gate}}\approx 0.9992$.
The robustness of FGQC against global control error was analytically demonstrated using perturbation method.
By comparing the numerical results of FGQC with NGQC and DG, the superiority of FGQC in solving global control error was confirmed.
Because FGQC is based on two-level systems, compared with NHQC, it has the advantage of not requiring complex quantum control on a multilevel structure. 
Therefore, this study makes a step towards the experimental realization of error-resilient quantum computation and quantum information processing.

\begin{acknowledgments}
This work was supported by the Natural Science Foundation of Guangdong Province (Grant No.2017B030308003), 
the Key R\&D Program of Guangdong province (Grant No. 2018B030326001), 
the Science, Technology and Innovation Commission of Shenzhen Municipality (Grant No.JCYJ20170412152620376 and No.JCYJ20170817105046702 and No.KYTDPT20181011104202253), 
the National Natural Science Foundation of China (Grant No.11875160, No.U1801661 and No.11905100),
the Economy, Trade and Information Commission of Shenzhen Municipality (Grant No.201901161512), 
the Guangdong Provincial Key Laboratory(Grant No.2019B121203002).
\end{acknowledgments}

\appendix

\section{\label{sec:detail}The detail of other scenarios}
The detail of other scenarios in simulating the Z gate. 
\textit{The NGQC Z gate}. The corresponding Hamiltonian reads
\begin{align}
H( t) & =(\Omega _{0}/2) e^{-i\phi ( t)} |1\rangle \langle 0|+\text{H.c.},
\label{eq:hxy}
\end{align}
where 
\begin{align}
\phi ( t) & =\begin{cases}
\pi  & 0\leq t< \tau _{\text{N1Z}},\\
-\pi /2 & \tau _{\text{N1Z}} \leq t\leq \tau _{\text{NZ}},
\end{cases}
\end{align}
with $\displaystyle \tau _{\text{N1Z}} =\tau_{\text{NZ}}/2$ and the run time $\displaystyle \tau _{\text{NZ}} =2\pi /\Omega _{0} = 0.5\mu \text{s}$. 
\textit{The standard dynamical Z gate}. For this scenario, the Hamiltonian can be written in the same form as Eq. (14) in the main text 
, but with $\Delta(t)=\Omega_{0}$, $\epsilon(t)=0$ and the run time $\tau_{\text{DSZ}}=\pi/\Omega_{0}=0.25\mu\text{s}$.

The detail of other scenarios in simulating the X gate. 
\textit{The NGQC X gate}. The Hamiltonian of this scenario can be written as the same form as Eq. \eqref{eq:hxy} but with
\begin{align}
   \phi(t)& =\begin{cases}
-\pi/2  & 0\leq t< \tau _{\text{NX1}},\\
\pi & \tau _{\text{NX1}} \leq t< \tau _{\text{NX2}}, \\
-\pi/2 & \tau _{\text{NX2}} \leq t\leq \tau _{\text{NX}},
\end{cases} 
\end{align}
where $\tau_{\text{NX1}}=(1/4)\tau_{\text{NX}}$, $\tau_{\text{NX2}}=(3/4)\tau_{\text{NX}}$ and $\tau_{\text{NX}}=2\pi/\Omega_{0}=0.5\mu\text{s}$.
\textit{The standard dynamical X gate.} The Hamiltonian of this scenario can be written in the same form as Eq. (14) in the main text 
with $\Delta(t)=0$, $\varphi(t)=0$, we set $\epsilon(t)=\Omega_{0}$ and the run time $\tau_{\text{DSX}}=\pi/\Omega_{0}=0.25\mu\text{s}$.

The detail of the NGQC two-qubit gate. The corresponding Hamiltonian of the NGQC two-qubit gate \cite{Zhao2017} can be written in the form of Eq. (23) in the main text 
, but $\Omega_{R}(t)$ is time-independent. We set: $\Omega_{R}(t)=\Omega_{0}$, the run time $\tau=2\pi/\Omega_{0}=0.5$ $\mu$s, other parameters are the same as the FGQC gate.

\section{\label{sec:T_H}Two more gates: the T and Hadamard gates}
\begin{figure}
\centering
\includegraphics[width=0.43\textwidth]{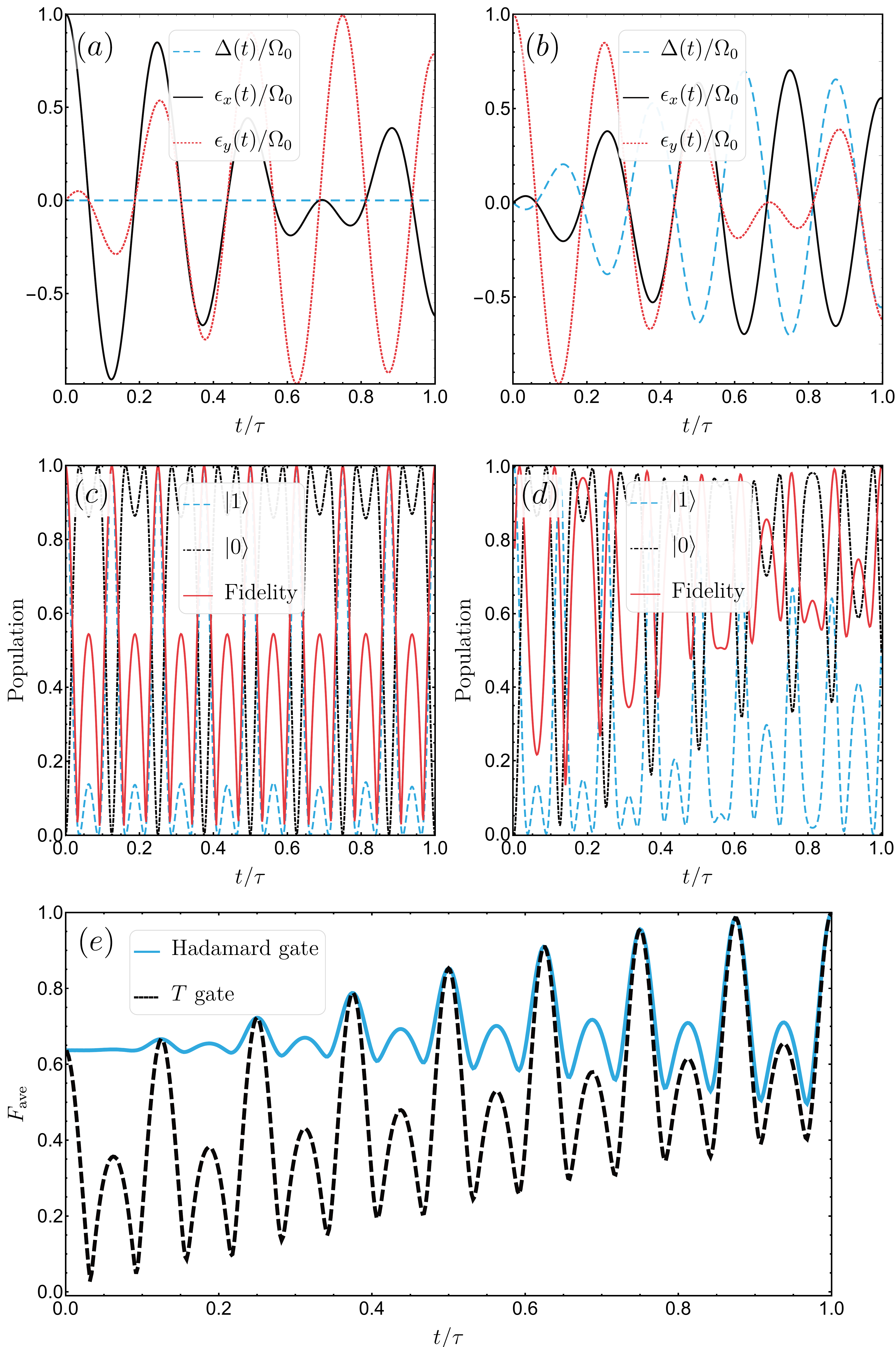}
\caption{The pulse shapes in numerically simulating the Hadamard (a) and T gates(b). 
The temporal evolution of populations (blue dashe line for state $|1\rangle$, black dotdashed line for state $|0\rangle$) and fidelities (red solid line)  with a given initial state $|\psi(0)\rangle=|1\rangle$ for the FGQC Hadamard gate (c) and T gates (d). 
(e) The temporal evolution of gate fidelities for the Hadamard (blue dashed line) and T gates (black solid line).}
\label{fig:h_and_t}
\end{figure}

To illustrate the universality of the FGQC, we numerically simulated the T and Hadamard gates using FGQC theory. For both the T and Hadamard gates, the Hamiltonian can be written in the following form
\begin{align}
    H(t)=\frac{\Delta(t)}{2}\sigma_{z}+\frac{\epsilon_{x}(t)}{2}\sigma_{x}+\frac{\epsilon_{y}(t)}{2}\sigma_{y}.
    \label{eq:h_ham}
\end{align}
For the FGQC Hadamard gate: $\Delta(t)=-\Omega_{0}\cos(\omega t)\sin(N_{H}t)/\sqrt{2}$, $\epsilon _{x}( t) =\Omega _{0}\cos( \omega t)\sin( N_{H} t) /\sqrt{2}$ and $\epsilon _{y}( t) =\Omega _{0}\cos( \omega t)\cos( N_{H} t)$ with $\Omega _{0} = 2.0\times 2\pi $ MHz, $\omega \approx 0.513\times 2\pi$ MHz $N_{H} \approx 45.728\times 2\pi $ KHz and the run time $\tau _{H} \approx 7.797\ \mu \text{s}$. This kind of Hamiltonian can be obtained through applying a rotating frame transformation: $H_{0}(t)\rightarrow H(t)= U_{0}^{\dagger}H_{0}(t)U_{0}(t)$ to $H_{0}(t)=\cos(\omega t)[\sin(N_{H}t)\sigma_{x}+\cos(N_{H}t)\sigma_{y}]$ with $U_{0}=\exp[-i(\pi/4)\sigma_{y}/2]$. For the FGQC T gate $\Delta(t) =0$, $\epsilon_{x}(t)=\Omega_{0}\cos(\omega t)\cos(N_{T}t)$, $\epsilon _{y}( t) =\Omega _{0}\cos( \omega t)\sin( N_{T}t)$ with $\Omega _{0} = 2.0\times 2\pi $ MHz, $\omega \approx 0.513\times 2\pi $ MHz $N_{T} \approx 45.728\times 2\pi $ KHz and the run time $\tau = 1.95\ \mu \text{s}$. The Hamiltonian of the FGQC T and Z gate has the same form, and the parameters of these two gates are almost the same except the run time of the T gate is shorter. The pulse shapes of the Hadamard and T gates are shown in Fig.\ref{fig:h_and_t} (a) and (b) respectively.  In Fig. \ref{fig:h_and_t} (c) and Fig. \ref{fig:h_and_t} (d), for the Hadamard and the T gate respectively, 
we show the we show the temporal evolution of state populations for $|0\rangle$ and $|1\rangle$ with a given initial state $|\psi(0)\rangle=|1\rangle$.
In the presence of the global control error ($\delta=0.1$) and the decoherence ($\gamma_{1}=8$Hz, $\gamma_{2}=80$ Hz), the corresponding fidelities between the target states and the temporal evolution states are also shown in these figures (red solid lines), at the end of the run time, $F_{H}(\tau)\approx 0.999338$ and $F_{T}(\tau)\approx 0.999263$. These fairly high fidelities of noisy gates provide evidence for the possibility of FGQC scenarios in real experiment. 
We have also investigated the gate fidelity of the FGQC Hadamard and T gates defined by $F=(1/2\pi)\int_{0}^{2\pi}\langle\psi_{I}|\rho|\psi_{I}\rangle d\Theta$ for initial states of the form $|\psi\rangle=\cos\Theta |0\rangle+\sin\Theta |1\rangle$, where a total of 101 different values of $\Theta$ were uniformly chosen in the range $[0,2\pi]$, the results are shown in Fig. \ref{fig:h_and_t} (e), the average fidelities at $\tau$ for the Hadamard and T gates are approximately $0.999338$ and $0.99984$ respectively.

\begin{figure}
    \centering
    \includegraphics[width=0.43\textwidth]{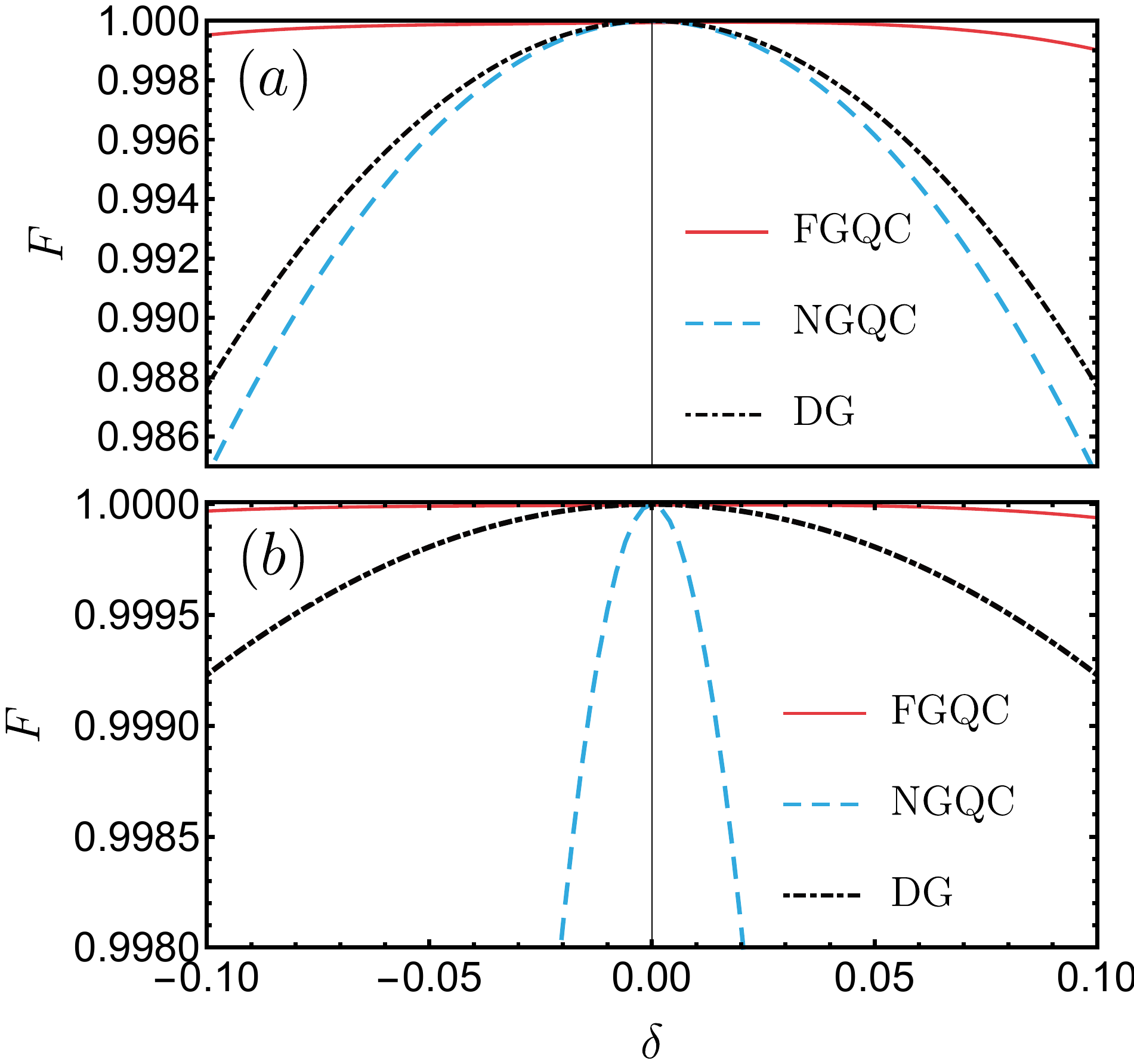}
    \caption{Fidelities of the Hadamard (a) and T (b) gates versus the amplitude of the global control error (without considering the decoherence in the simulation),  results of four different protocols are shown:  FGQC (red solid line), NGQC (blue dashed line) and standard DG (black dotdashed line).}
    \label{fig:f_versus_error}
\end{figure}

In terms of the robustness against the global control error, we compare the performance of the FGQC Hadamard and T gates with those of other typical scenarios based on two-level systems: the nonadiabatic geometric quantum computation (NGQC), the standard dynamical gates (DG). The results are shown in Fig. \ref{fig:f_versus_error}. 
It is readily to see from these figures that the FGQC scenarios perform much better than other three scenarios.

The detail of other scenarios in simulating the T gate. 
We set $\Omega_{0}=2\times 2\pi$ MHz hereinafter.
\textit{The NGQC T gate}: The  Hamiltonian of this scenario can be written as 
\begin{align}
    H_{\text{NT}}=(\Omega_{0}/2)e^{-i\phi(t)}|1\rangle\langle 0|+\text{H.c.}, 
    \label{eq:ngqct}
\end{align}
where 
\begin{align}
\phi ( t) & =\begin{cases}
\pi /2 & 0\leqslant t< \tau _{\text{NT}} /2,\\
\pi /2+\pi /8 & \tau _{\text{NT}} /2\leqslant t\leqslant \tau _{\text{NT}},
\end{cases}
\end{align}
the run time $\tau _{\text{NT}} =2\pi /\Omega _{0} = 0.5\mu \text{s}$. 
\textit{The standard dynamical T gate}: the corresponding Hamiltonian reads $H(t)=\Omega_{0}\sigma_{z}/2$ and the run time is given by $\tau_{\text{DGT}}=(\pi/4)/\Omega_{0}= 0.0625\mu\text{s}$. 

The detail of other scenarios in simulating the Hadamard gate. \textit{The NGQC Hadamard gate}: the Hamiltonian has the same form as Eq. \eqref{eq:ngqct}, but the time-dependent phase $\phi(t)$ is given by the following expression: 
\begin{align}
\phi ( t) & =\begin{cases}
-\pi /2 & 0\leqslant t< \tau _{\text{NH1}},\\
0 & \tau _{\text{NH1}} \leqslant t <\tau _{\text{NH2}},\\
-\pi /2 & \tau _{\text{NH2}} \leqslant t\leqslant \tau _{\text{NH}},
\end{cases}
\end{align}
with $\tau _{\text{NH1}} =\tau _{\text{NH}} /8$, $\tau _{\text{NH2}} =5\tau _{\text{NH}} /8$, $\tau _{\text{NH}} =2\pi /\Omega _{0} = 0.5\mu \text{s}$.
\textit{The standard dynamical Hadamard gate}: the Hamiltonian reads $H(t)=\Omega_{0}(\sigma_{x}+\sigma_{z})/2\sqrt{2}$, the run time $\tau_{\text{DGH}}=\pi/\Omega_{0}=0.25\mu$s.

\bibliography{manuscript}

\begin{thebibliography}{90}%
\makeatletter
\providecommand \@ifxundefined [1]{%
 \@ifx{#1\undefined}
}%
\providecommand \@ifnum [1]{%
 \ifnum #1\expandafter \@firstoftwo
 \else \expandafter \@secondoftwo
 \fi
}%
\providecommand \@ifx [1]{%
 \ifx #1\expandafter \@firstoftwo
 \else \expandafter \@secondoftwo
 \fi
}%
\providecommand \natexlab [1]{#1}%
\providecommand \enquote  [1]{``#1''}%
\providecommand \bibnamefont  [1]{#1}%
\providecommand \bibfnamefont [1]{#1}%
\providecommand \citenamefont [1]{#1}%
\providecommand \href@noop [0]{\@secondoftwo}%
\providecommand \href [0]{\begingroup \@sanitize@url \@href}%
\providecommand \@href[1]{\@@startlink{#1}\@@href}%
\providecommand \@@href[1]{\endgroup#1\@@endlink}%
\providecommand \@sanitize@url [0]{\catcode `\\12\catcode `\$12\catcode
  `\&12\catcode `\#12\catcode `\^12\catcode `\_12\catcode `\%12\relax}%
\providecommand \@@startlink[1]{}%
\providecommand \@@endlink[0]{}%
\providecommand \url  [0]{\begingroup\@sanitize@url \@url }%
\providecommand \@url [1]{\endgroup\@href {#1}{\urlprefix }}%
\providecommand \urlprefix  [0]{URL }%
\providecommand \Eprint [0]{\href }%
\providecommand \doibase [0]{http://dx.doi.org/}%
\providecommand \selectlanguage [0]{\@gobble}%
\providecommand \bibinfo  [0]{\@secondoftwo}%
\providecommand \bibfield  [0]{\@secondoftwo}%
\providecommand \translation [1]{[#1]}%
\providecommand \BibitemOpen [0]{}%
\providecommand \bibitemStop [0]{}%
\providecommand \bibitemNoStop [0]{.\EOS\space}%
\providecommand \EOS [0]{\spacefactor3000\relax}%
\providecommand \BibitemShut  [1]{\csname bibitem#1\endcsname}%
\let\auto@bib@innerbib\@empty
\bibitem [{\citenamefont {Feynman}(1999)}]{feynman1999}%
  \BibitemOpen
  \bibfield  {author} {\bibinfo {author} {\bibfnamefont {Richard~P}\
  \bibnamefont {Feynman}},\ }\bibfield  {title} {\enquote {\bibinfo {title}
  {Simulating physics with computers},}\ }\href@noop {} {\bibfield  {journal}
  {\bibinfo  {journal} {Int. J. Theor. Phys}\ }\textbf {\bibinfo {volume} {21}}
  (\bibinfo {year} {1999})}\BibitemShut {NoStop}%
\bibitem [{\citenamefont {Li}\ \emph {et~al.}(2014)\citenamefont {Li},
  \citenamefont {Zhou}, \citenamefont {Ju}, \citenamefont {Chen}, \citenamefont
  {Zheng}, \citenamefont {Lu}, \citenamefont {Rong}, \citenamefont {Duan},
  \citenamefont {Peng},\ and\ \citenamefont {Du}}]{PhysRevLett.112.220501}%
  \BibitemOpen
  \bibfield  {author} {\bibinfo {author} {\bibfnamefont {Zhaokai}\ \bibnamefont
  {Li}}, \bibinfo {author} {\bibfnamefont {Hui}\ \bibnamefont {Zhou}}, \bibinfo
  {author} {\bibfnamefont {Chenyong}\ \bibnamefont {Ju}}, \bibinfo {author}
  {\bibfnamefont {Hongwei}\ \bibnamefont {Chen}}, \bibinfo {author}
  {\bibfnamefont {Wenqiang}\ \bibnamefont {Zheng}}, \bibinfo {author}
  {\bibfnamefont {Dawei}\ \bibnamefont {Lu}}, \bibinfo {author} {\bibfnamefont
  {Xing}\ \bibnamefont {Rong}}, \bibinfo {author} {\bibfnamefont {Changkui}\
  \bibnamefont {Duan}}, \bibinfo {author} {\bibfnamefont {Xinhua}\ \bibnamefont
  {Peng}}, \ and\ \bibinfo {author} {\bibfnamefont {Jiangfeng}\ \bibnamefont
  {Du}},\ }\bibfield  {title} {\enquote {\bibinfo {title} {Experimental
  realization of a compressed quantum simulation of a 32-spin ising chain},}\
  }\href {\doibase 10.1103/PhysRevLett.112.220501} {\bibfield  {journal}
  {\bibinfo  {journal} {Phys. Rev. Lett.}\ }\textbf {\bibinfo {volume} {112}},\
  \bibinfo {pages} {220501} (\bibinfo {year} {2014})}\BibitemShut {NoStop}%
\bibitem [{\citenamefont {Vandersypen}\ \emph {et~al.}(2001)\citenamefont
  {Vandersypen}, \citenamefont {Steffen}, \citenamefont {Breyta}, \citenamefont
  {Yannoni}, \citenamefont {Sherwood},\ and\ \citenamefont
  {Chuang}}]{Vandersypen2001}%
  \BibitemOpen
  \bibfield  {author} {\bibinfo {author} {\bibfnamefont {Lieven M.~K.}\
  \bibnamefont {Vandersypen}}, \bibinfo {author} {\bibfnamefont {Matthias}\
  \bibnamefont {Steffen}}, \bibinfo {author} {\bibfnamefont {Gregory}\
  \bibnamefont {Breyta}}, \bibinfo {author} {\bibfnamefont {Costantino~S.}\
  \bibnamefont {Yannoni}}, \bibinfo {author} {\bibfnamefont {Mark~H.}\
  \bibnamefont {Sherwood}}, \ and\ \bibinfo {author} {\bibfnamefont {Isaac~L.}\
  \bibnamefont {Chuang}},\ }\bibfield  {title} {\enquote {\bibinfo {title}
  {Experimental realization of shor's quantum factoring algorithm using nuclear
  magnetic resonance},}\ }\href {https://doi.org/10.1038/414883a} {\bibfield
  {journal} {\bibinfo  {journal} {Nature}\ }\textbf {\bibinfo {volume} {414}},\
  \bibinfo {pages} {883--887} (\bibinfo {year} {2001})}\BibitemShut {NoStop}%
\bibitem [{\citenamefont {Xu}\ \emph {et~al.}(2012{\natexlab{a}})\citenamefont
  {Xu}, \citenamefont {Zhu}, \citenamefont {Lu}, \citenamefont {Zhou},
  \citenamefont {Peng},\ and\ \citenamefont {Du}}]{Xu2012}%
  \BibitemOpen
  \bibfield  {author} {\bibinfo {author} {\bibfnamefont {Nanyang}\ \bibnamefont
  {Xu}}, \bibinfo {author} {\bibfnamefont {Jing}\ \bibnamefont {Zhu}}, \bibinfo
  {author} {\bibfnamefont {Dawei}\ \bibnamefont {Lu}}, \bibinfo {author}
  {\bibfnamefont {Xianyi}\ \bibnamefont {Zhou}}, \bibinfo {author}
  {\bibfnamefont {Xinhua}\ \bibnamefont {Peng}}, \ and\ \bibinfo {author}
  {\bibfnamefont {Jiangfeng}\ \bibnamefont {Du}},\ }\bibfield  {title}
  {\enquote {\bibinfo {title} {Quantum factorization of 143 on a
  dipolar-coupling nuclear magnetic resonance system},}\ }\href {\doibase
  10.1103/PhysRevLett.108.130501} {\bibfield  {journal} {\bibinfo  {journal}
  {Phys. Rev. Lett.}\ }\textbf {\bibinfo {volume} {108}},\ \bibinfo {pages}
  {130501} (\bibinfo {year} {2012}{\natexlab{a}})}\BibitemShut {NoStop}%
\bibitem [{\citenamefont {Martín-López}\ \emph {et~al.}(2012)\citenamefont
  {Martín-López}, \citenamefont {Laing}, \citenamefont {Lawson},
  \citenamefont {Alvarez}, \citenamefont {Zhou},\ and\ \citenamefont
  {O'Brien}}]{Martin-Lopez2012}%
  \BibitemOpen
  \bibfield  {author} {\bibinfo {author} {\bibfnamefont {Enrique}\ \bibnamefont
  {Martín-López}}, \bibinfo {author} {\bibfnamefont {Anthony}\ \bibnamefont
  {Laing}}, \bibinfo {author} {\bibfnamefont {Thomas}\ \bibnamefont {Lawson}},
  \bibinfo {author} {\bibfnamefont {Roberto}\ \bibnamefont {Alvarez}}, \bibinfo
  {author} {\bibfnamefont {Xiao-Qi}\ \bibnamefont {Zhou}}, \ and\ \bibinfo
  {author} {\bibfnamefont {Jeremy~L.}\ \bibnamefont {O'Brien}},\ }\bibfield
  {title} {\enquote {\bibinfo {title} {Experimental realization of shor's
  quantum factoring algorithm using qubit recycling},}\ }\href
  {https://doi.org/10.1038/nphoton.2012.259} {\bibfield  {journal} {\bibinfo
  {journal} {Nature Photonics}\ }\textbf {\bibinfo {volume} {6}},\ \bibinfo
  {pages} {773--776} (\bibinfo {year} {2012})}\BibitemShut {NoStop}%
\bibitem [{\citenamefont {Grover}(1997)}]{Grover1997}%
  \BibitemOpen
  \bibfield  {author} {\bibinfo {author} {\bibfnamefont {Lov~K.}\ \bibnamefont
  {Grover}},\ }\bibfield  {title} {\enquote {\bibinfo {title} {Quantum
  mechanics helps in searching for a needle in a haystack},}\ }\href {\doibase
  10.1103/PhysRevLett.79.325} {\bibfield  {journal} {\bibinfo  {journal} {Phys.
  Rev. Lett.}\ }\textbf {\bibinfo {volume} {79}},\ \bibinfo {pages} {325--328}
  (\bibinfo {year} {1997})}\BibitemShut {NoStop}%
\bibitem [{\citenamefont {Rebentrost}\ \emph {et~al.}(2014)\citenamefont
  {Rebentrost}, \citenamefont {Mohseni},\ and\ \citenamefont
  {Lloyd}}]{Rebentrost2014}%
  \BibitemOpen
  \bibfield  {author} {\bibinfo {author} {\bibfnamefont {Patrick}\ \bibnamefont
  {Rebentrost}}, \bibinfo {author} {\bibfnamefont {Masoud}\ \bibnamefont
  {Mohseni}}, \ and\ \bibinfo {author} {\bibfnamefont {Seth}\ \bibnamefont
  {Lloyd}},\ }\bibfield  {title} {\enquote {\bibinfo {title} {Quantum support
  vector machine for big data classification},}\ }\href {\doibase
  10.1103/PhysRevLett.113.130503} {\bibfield  {journal} {\bibinfo  {journal}
  {Phys. Rev. Lett.}\ }\textbf {\bibinfo {volume} {113}},\ \bibinfo {pages}
  {130503} (\bibinfo {year} {2014})}\BibitemShut {NoStop}%
\bibitem [{\citenamefont {Li}\ \emph {et~al.}(2015)\citenamefont {Li},
  \citenamefont {Liu}, \citenamefont {Xu},\ and\ \citenamefont
  {Du}}]{PhysRevLett.114.140504}%
  \BibitemOpen
  \bibfield  {author} {\bibinfo {author} {\bibfnamefont {Zhaokai}\ \bibnamefont
  {Li}}, \bibinfo {author} {\bibfnamefont {Xiaomei}\ \bibnamefont {Liu}},
  \bibinfo {author} {\bibfnamefont {Nanyang}\ \bibnamefont {Xu}}, \ and\
  \bibinfo {author} {\bibfnamefont {Jiangfeng}\ \bibnamefont {Du}},\ }\bibfield
   {title} {\enquote {\bibinfo {title} {Experimental realization of a quantum
  support vector machine},}\ }\href {\doibase 10.1103/PhysRevLett.114.140504}
  {\bibfield  {journal} {\bibinfo  {journal} {Phys. Rev. Lett.}\ }\textbf
  {\bibinfo {volume} {114}},\ \bibinfo {pages} {140504} (\bibinfo {year}
  {2015})}\BibitemShut {NoStop}%
\bibitem [{\citenamefont {Cong}\ \emph {et~al.}(2019)\citenamefont {Cong},
  \citenamefont {Choi},\ and\ \citenamefont {Lukin}}]{Cong2019}%
  \BibitemOpen
  \bibfield  {author} {\bibinfo {author} {\bibfnamefont {Iris}\ \bibnamefont
  {Cong}}, \bibinfo {author} {\bibfnamefont {Soonwon}\ \bibnamefont {Choi}}, \
  and\ \bibinfo {author} {\bibfnamefont {Mikhail~D.}\ \bibnamefont {Lukin}},\
  }\bibfield  {title} {\enquote {\bibinfo {title} {Quantum convolutional neural
  networks},}\ }\href {https://doi.org/10.1038/s41567-019-0648-8} {\bibfield
  {journal} {\bibinfo  {journal} {Nature Physics}\ }\textbf {\bibinfo {volume}
  {15}},\ \bibinfo {pages} {1273--1278} (\bibinfo {year} {2019})}\BibitemShut
  {NoStop}%
\bibitem [{\citenamefont {Berry}(1984)}]{berry1984quantal}%
  \BibitemOpen
  \bibfield  {author} {\bibinfo {author} {\bibfnamefont {Michael~Victor}\
  \bibnamefont {Berry}},\ }\bibfield  {title} {\enquote {\bibinfo {title}
  {Quantal phase factors accompanying adiabatic changes},}\ }\href
  {http://doi.org/10.1098/rspa.1984.0023} {\bibfield  {journal} {\bibinfo
  {journal} {Proc. R. Soc. Lond. A.}\ }\textbf {\bibinfo {volume} {392}},\
  \bibinfo {pages} {45--57} (\bibinfo {year} {1984})}\BibitemShut {NoStop}%
\bibitem [{\citenamefont {Aharonov}\ and\ \citenamefont
  {Anandan}(1987)}]{Aharonov1987}%
  \BibitemOpen
  \bibfield  {author} {\bibinfo {author} {\bibfnamefont {Y.}~\bibnamefont
  {Aharonov}}\ and\ \bibinfo {author} {\bibfnamefont {J.}~\bibnamefont
  {Anandan}},\ }\bibfield  {title} {\enquote {\bibinfo {title} {Phase change
  during a cyclic quantum evolution},}\ }\href {\doibase
  10.1103/PhysRevLett.58.1593} {\bibfield  {journal} {\bibinfo  {journal}
  {Phys. Rev. Lett.}\ }\textbf {\bibinfo {volume} {58}},\ \bibinfo {pages}
  {1593--1596} (\bibinfo {year} {1987})}\BibitemShut {NoStop}%
\bibitem [{\citenamefont {Wilczek}\ and\ \citenamefont
  {Zee}(1984)}]{Wilczek1984}%
  \BibitemOpen
  \bibfield  {author} {\bibinfo {author} {\bibfnamefont {Frank}\ \bibnamefont
  {Wilczek}}\ and\ \bibinfo {author} {\bibfnamefont {A.}~\bibnamefont {Zee}},\
  }\bibfield  {title} {\enquote {\bibinfo {title} {Appearance of gauge
  structure in simple dynamical systems},}\ }\href {\doibase
  10.1103/PhysRevLett.52.2111} {\bibfield  {journal} {\bibinfo  {journal}
  {Phys. Rev. Lett.}\ }\textbf {\bibinfo {volume} {52}},\ \bibinfo {pages}
  {2111--2114} (\bibinfo {year} {1984})}\BibitemShut {NoStop}%
\bibitem [{\citenamefont {Anandan}(1988)}]{ANANDAN1988171}%
  \BibitemOpen
  \bibfield  {author} {\bibinfo {author} {\bibfnamefont {J.}~\bibnamefont
  {Anandan}},\ }\bibfield  {title} {\enquote {\bibinfo {title} {Non-adiabatic
  non-abelian geometric phase},}\ }\href {\doibase
  https://doi.org/10.1016/0375-9601(88)91010-9} {\bibfield  {journal} {\bibinfo
   {journal} {Physics Letters A}\ }\textbf {\bibinfo {volume} {133}},\ \bibinfo
  {pages} {171 -- 175} (\bibinfo {year} {1988})}\BibitemShut {NoStop}%
\bibitem [{\citenamefont {Chiara}\ and\ \citenamefont
  {Palma}(2003)}]{Chiara2003Berry}%
  \BibitemOpen
  \bibfield  {author} {\bibinfo {author} {\bibfnamefont {Gabriele~De}\
  \bibnamefont {Chiara}}\ and\ \bibinfo {author} {\bibfnamefont {G.~Massimo}\
  \bibnamefont {Palma}},\ }\bibfield  {title} {\enquote {\bibinfo {title}
  {Berry phase for a spin1/2particle in a classical fluctuating field},}\
  }\href@noop {} {\bibfield  {journal} {\bibinfo  {journal} {Physical Review
  Letters}\ } (\bibinfo {year} {2003})}\BibitemShut {NoStop}%
\bibitem [{\citenamefont {Shiliang}\ and\ \citenamefont
  {Zanardi}(2005)}]{2005Geometric}%
  \BibitemOpen
  \bibfield  {author} {\bibinfo {author} {\bibfnamefont {Zhu}\ \bibnamefont
  {Shiliang}}\ and\ \bibinfo {author} {\bibfnamefont {Paolo}\ \bibnamefont
  {Zanardi}},\ }\bibfield  {title} {\enquote {\bibinfo {title} {Geometric
  quantum gates that are robust against stochastic control errors},}\
  }\href@noop {} {\bibfield  {journal} {\bibinfo  {journal} {Physical Review
  A}\ }\textbf {\bibinfo {volume} {72}},\ \bibinfo {pages} {656--665} (\bibinfo
  {year} {2005})}\BibitemShut {NoStop}%
\bibitem [{\citenamefont {Leek}\ \emph {et~al.}(2007)\citenamefont {Leek},
  \citenamefont {Fink}, \citenamefont {Blais}, \citenamefont {Bianchetti},
  \citenamefont {G{\"o}ppl}, \citenamefont {Gambetta}, \citenamefont
  {Schuster}, \citenamefont {Frunzio}, \citenamefont {Schoelkopf},\ and\
  \citenamefont {Wallraff}}]{Leek1889}%
  \BibitemOpen
  \bibfield  {author} {\bibinfo {author} {\bibfnamefont {P.~J.}\ \bibnamefont
  {Leek}}, \bibinfo {author} {\bibfnamefont {J.~M.}\ \bibnamefont {Fink}},
  \bibinfo {author} {\bibfnamefont {A.}~\bibnamefont {Blais}}, \bibinfo
  {author} {\bibfnamefont {R.}~\bibnamefont {Bianchetti}}, \bibinfo {author}
  {\bibfnamefont {M.}~\bibnamefont {G{\"o}ppl}}, \bibinfo {author}
  {\bibfnamefont {J.~M.}\ \bibnamefont {Gambetta}}, \bibinfo {author}
  {\bibfnamefont {D.~I.}\ \bibnamefont {Schuster}}, \bibinfo {author}
  {\bibfnamefont {L.}~\bibnamefont {Frunzio}}, \bibinfo {author} {\bibfnamefont
  {R.~J.}\ \bibnamefont {Schoelkopf}}, \ and\ \bibinfo {author} {\bibfnamefont
  {A.}~\bibnamefont {Wallraff}},\ }\bibfield  {title} {\enquote {\bibinfo
  {title} {Observation of berry{\textquoteright}s phase in a solid-state
  qubit},}\ }\href {\doibase 10.1126/science.1149858} {\bibfield  {journal}
  {\bibinfo  {journal} {Science}\ }\textbf {\bibinfo {volume} {318}},\ \bibinfo
  {pages} {1889--1892} (\bibinfo {year} {2007})}\BibitemShut {NoStop}%
\bibitem [{\citenamefont {Filipp}\ \emph {et~al.}(2009)\citenamefont {Filipp},
  \citenamefont {Klepp}, \citenamefont {Hasegawa}, \citenamefont
  {Plonka-Spehr}, \citenamefont {Schmidt}, \citenamefont {Geltenbort},\ and\
  \citenamefont {Rauch}}]{2009Experimental}%
  \BibitemOpen
  \bibfield  {author} {\bibinfo {author} {\bibfnamefont {S.}~\bibnamefont
  {Filipp}}, \bibinfo {author} {\bibfnamefont {J.}~\bibnamefont {Klepp}},
  \bibinfo {author} {\bibfnamefont {Y.}~\bibnamefont {Hasegawa}}, \bibinfo
  {author} {\bibfnamefont {C.}~\bibnamefont {Plonka-Spehr}}, \bibinfo {author}
  {\bibfnamefont {U.}~\bibnamefont {Schmidt}}, \bibinfo {author} {\bibfnamefont
  {P.}~\bibnamefont {Geltenbort}}, \ and\ \bibinfo {author} {\bibfnamefont
  {H.}~\bibnamefont {Rauch}},\ }\bibfield  {title} {\enquote {\bibinfo {title}
  {Experimental demonstration of the stability of berry's phase for a spin-1/2
  particle},}\ }\href@noop {} {\bibfield  {journal} {\bibinfo  {journal}
  {Physical Review Letters}\ }\textbf {\bibinfo {volume} {102}},\ \bibinfo
  {pages} {404} (\bibinfo {year} {2009})}\BibitemShut {NoStop}%
\bibitem [{\citenamefont {Berger}\ \emph {et~al.}(2013)\citenamefont {Berger},
  \citenamefont {Pechal}, \citenamefont {Jr}, \citenamefont {Eichler},
  \citenamefont {Steffen}, \citenamefont {Fedorov}, \citenamefont {Wallraff},\
  and\ \citenamefont {Filipp}}]{2013Exploring}%
  \BibitemOpen
  \bibfield  {author} {\bibinfo {author} {\bibfnamefont {S.}~\bibnamefont
  {Berger}}, \bibinfo {author} {\bibfnamefont {M.}~\bibnamefont {Pechal}},
  \bibinfo {author} {\bibfnamefont {A.~A.~Abdumalikov}\ \bibnamefont {Jr}},
  \bibinfo {author} {\bibfnamefont {C.}~\bibnamefont {Eichler}}, \bibinfo
  {author} {\bibfnamefont {L.}~\bibnamefont {Steffen}}, \bibinfo {author}
  {\bibfnamefont {A.}~\bibnamefont {Fedorov}}, \bibinfo {author} {\bibfnamefont
  {A.}~\bibnamefont {Wallraff}}, \ and\ \bibinfo {author} {\bibfnamefont
  {S.}~\bibnamefont {Filipp}},\ }\bibfield  {title} {\enquote {\bibinfo {title}
  {Exploring the effect of noise on the berry phase},}\ }\href@noop {}
  {\bibfield  {journal} {\bibinfo  {journal} {Physical Review A}\ }\textbf
  {\bibinfo {volume} {87}},\ \bibinfo {pages} {1--5} (\bibinfo {year}
  {2013})}\BibitemShut {NoStop}%
\bibitem [{\citenamefont {Jones}\ \emph {et~al.}(2000)\citenamefont {Jones},
  \citenamefont {Vedral}, \citenamefont {Ekert},\ and\ \citenamefont
  {Castagnoli}}]{Jones2000}%
  \BibitemOpen
  \bibfield  {author} {\bibinfo {author} {\bibfnamefont {Jonathan~A.}\
  \bibnamefont {Jones}}, \bibinfo {author} {\bibfnamefont {Vlatko}\
  \bibnamefont {Vedral}}, \bibinfo {author} {\bibfnamefont {Artur}\
  \bibnamefont {Ekert}}, \ and\ \bibinfo {author} {\bibfnamefont {Giuseppe}\
  \bibnamefont {Castagnoli}},\ }\bibfield  {title} {\enquote {\bibinfo {title}
  {Geometric quantum computation using nuclear magnetic resonance},}\ }\href
  {https://doi.org/10.1038/35002528} {\bibfield  {journal} {\bibinfo  {journal}
  {Nature}\ }\textbf {\bibinfo {volume} {403}},\ \bibinfo {pages} {869--871}
  (\bibinfo {year} {2000})}\BibitemShut {NoStop}%
\bibitem [{\citenamefont {Duan}\ \emph {et~al.}(2001)\citenamefont {Duan},
  \citenamefont {Cirac},\ and\ \citenamefont {Zoller}}]{Duan1695}%
  \BibitemOpen
  \bibfield  {author} {\bibinfo {author} {\bibfnamefont {L.-M.}\ \bibnamefont
  {Duan}}, \bibinfo {author} {\bibfnamefont {J.~I.}\ \bibnamefont {Cirac}}, \
  and\ \bibinfo {author} {\bibfnamefont {P.}~\bibnamefont {Zoller}},\
  }\bibfield  {title} {\enquote {\bibinfo {title} {Geometric manipulation of
  trapped ions for quantum computation},}\ }\href {\doibase
  10.1126/science.1058835} {\bibfield  {journal} {\bibinfo  {journal}
  {Science}\ }\textbf {\bibinfo {volume} {292}},\ \bibinfo {pages} {1695--1697}
  (\bibinfo {year} {2001})}\BibitemShut {NoStop}%
\bibitem [{\citenamefont {Wu}\ \emph {et~al.}(2005)\citenamefont {Wu},
  \citenamefont {Zanardi},\ and\ \citenamefont {Lidar}}]{Wu2005}%
  \BibitemOpen
  \bibfield  {author} {\bibinfo {author} {\bibfnamefont {L.-A.}\ \bibnamefont
  {Wu}}, \bibinfo {author} {\bibfnamefont {P.}~\bibnamefont {Zanardi}}, \ and\
  \bibinfo {author} {\bibfnamefont {D.~A.}\ \bibnamefont {Lidar}},\ }\bibfield
  {title} {\enquote {\bibinfo {title} {Holonomic quantum computation in
  decoherence-free subspaces},}\ }\href {\doibase
  10.1103/PhysRevLett.95.130501} {\bibfield  {journal} {\bibinfo  {journal}
  {Phys. Rev. Lett.}\ }\textbf {\bibinfo {volume} {95}},\ \bibinfo {pages}
  {130501} (\bibinfo {year} {2005})}\BibitemShut {NoStop}%
\bibitem [{\citenamefont {Wu}\ \emph {et~al.}(2013)\citenamefont {Wu},
  \citenamefont {Gauger}, \citenamefont {George}, \citenamefont {M\"ott\"onen},
  \citenamefont {Riemann}, \citenamefont {Abrosimov}, \citenamefont {Becker},
  \citenamefont {Pohl}, \citenamefont {Itoh}, \citenamefont {Thewalt},\ and\
  \citenamefont {Morton}}]{Wu2013}%
  \BibitemOpen
  \bibfield  {author} {\bibinfo {author} {\bibfnamefont {Hua}\ \bibnamefont
  {Wu}}, \bibinfo {author} {\bibfnamefont {Erik~M.}\ \bibnamefont {Gauger}},
  \bibinfo {author} {\bibfnamefont {Richard~E.}\ \bibnamefont {George}},
  \bibinfo {author} {\bibfnamefont {Mikko}\ \bibnamefont {M\"ott\"onen}},
  \bibinfo {author} {\bibfnamefont {Helge}\ \bibnamefont {Riemann}}, \bibinfo
  {author} {\bibfnamefont {Nikolai~V.}\ \bibnamefont {Abrosimov}}, \bibinfo
  {author} {\bibfnamefont {Peter}\ \bibnamefont {Becker}}, \bibinfo {author}
  {\bibfnamefont {Hans-Joachim}\ \bibnamefont {Pohl}}, \bibinfo {author}
  {\bibfnamefont {Kohei~M.}\ \bibnamefont {Itoh}}, \bibinfo {author}
  {\bibfnamefont {Mike L.~W.}\ \bibnamefont {Thewalt}}, \ and\ \bibinfo
  {author} {\bibfnamefont {John J.~L.}\ \bibnamefont {Morton}},\ }\bibfield
  {title} {\enquote {\bibinfo {title} {Geometric phase gates with adiabatic
  control in electron spin resonance},}\ }\href {\doibase
  10.1103/PhysRevA.87.032326} {\bibfield  {journal} {\bibinfo  {journal} {Phys.
  Rev. A}\ }\textbf {\bibinfo {volume} {87}},\ \bibinfo {pages} {032326}
  (\bibinfo {year} {2013})}\BibitemShut {NoStop}%
\bibitem [{\citenamefont {Huang}\ \emph {et~al.}(2019)\citenamefont {Huang},
  \citenamefont {Wu}, \citenamefont {Wang}, \citenamefont {Hou}, \citenamefont
  {Wang}, \citenamefont {Zhang}, \citenamefont {Lian}, \citenamefont {Liu},
  \citenamefont {Wang}, \citenamefont {Zhang}, \citenamefont {He},
  \citenamefont {Chang}, \citenamefont {Xu},\ and\ \citenamefont
  {Duan}}]{Huang2019}%
  \BibitemOpen
  \bibfield  {author} {\bibinfo {author} {\bibfnamefont {Y.-Y.}\ \bibnamefont
  {Huang}}, \bibinfo {author} {\bibfnamefont {Y.-K.}\ \bibnamefont {Wu}},
  \bibinfo {author} {\bibfnamefont {F.}~\bibnamefont {Wang}}, \bibinfo {author}
  {\bibfnamefont {P.-Y.}\ \bibnamefont {Hou}}, \bibinfo {author} {\bibfnamefont
  {W.-B.}\ \bibnamefont {Wang}}, \bibinfo {author} {\bibfnamefont {W.-G.}\
  \bibnamefont {Zhang}}, \bibinfo {author} {\bibfnamefont {W.-Q.}\ \bibnamefont
  {Lian}}, \bibinfo {author} {\bibfnamefont {Y.-Q.}\ \bibnamefont {Liu}},
  \bibinfo {author} {\bibfnamefont {H.-Y.}\ \bibnamefont {Wang}}, \bibinfo
  {author} {\bibfnamefont {H.-Y.}\ \bibnamefont {Zhang}}, \bibinfo {author}
  {\bibfnamefont {L.}~\bibnamefont {He}}, \bibinfo {author} {\bibfnamefont
  {X.-Y.}\ \bibnamefont {Chang}}, \bibinfo {author} {\bibfnamefont
  {Y.}~\bibnamefont {Xu}}, \ and\ \bibinfo {author} {\bibfnamefont {L.-M.}\
  \bibnamefont {Duan}},\ }\bibfield  {title} {\enquote {\bibinfo {title}
  {Experimental realization of robust geometric quantum gates with solid-state
  spins},}\ }\href {\doibase 10.1103/PhysRevLett.122.010503} {\bibfield
  {journal} {\bibinfo  {journal} {Phys. Rev. Lett.}\ }\textbf {\bibinfo
  {volume} {122}},\ \bibinfo {pages} {010503} (\bibinfo {year}
  {2019})}\BibitemShut {NoStop}%
\bibitem [{\citenamefont {Xiang-Bin}\ and\ \citenamefont
  {Keiji}(2001)}]{Xiang-Bin2001}%
  \BibitemOpen
  \bibfield  {author} {\bibinfo {author} {\bibfnamefont {Wang}\ \bibnamefont
  {Xiang-Bin}}\ and\ \bibinfo {author} {\bibfnamefont {Matsumoto}\ \bibnamefont
  {Keiji}},\ }\bibfield  {title} {\enquote {\bibinfo {title} {Nonadiabatic
  conditional geometric phase shift with {NMR}},}\ }\href {\doibase
  10.1103/PhysRevLett.87.097901} {\bibfield  {journal} {\bibinfo  {journal}
  {Phys. Rev. Lett.}\ }\textbf {\bibinfo {volume} {87}},\ \bibinfo {pages}
  {097901} (\bibinfo {year} {2001})}\BibitemShut {NoStop}%
\bibitem [{\citenamefont {Zhu}\ and\ \citenamefont {Wang}(2002)}]{Zhu2002}%
  \BibitemOpen
  \bibfield  {author} {\bibinfo {author} {\bibfnamefont {Shi-Liang}\
  \bibnamefont {Zhu}}\ and\ \bibinfo {author} {\bibfnamefont {Z.~D.}\
  \bibnamefont {Wang}},\ }\bibfield  {title} {\enquote {\bibinfo {title}
  {Implementation of universal quantum gates based on nonadiabatic geometric
  phases},}\ }\href {\doibase 10.1103/PhysRevLett.89.097902} {\bibfield
  {journal} {\bibinfo  {journal} {Phys. Rev. Lett.}\ }\textbf {\bibinfo
  {volume} {89}},\ \bibinfo {pages} {097902} (\bibinfo {year}
  {2002})}\BibitemShut {NoStop}%
\bibitem [{\citenamefont {Thomas}\ \emph {et~al.}(2011)\citenamefont {Thomas},
  \citenamefont {Lababidi},\ and\ \citenamefont {Tian}}]{Thomas2011}%
  \BibitemOpen
  \bibfield  {author} {\bibinfo {author} {\bibfnamefont {J.~T.}\ \bibnamefont
  {Thomas}}, \bibinfo {author} {\bibfnamefont {Mahmoud}\ \bibnamefont
  {Lababidi}}, \ and\ \bibinfo {author} {\bibfnamefont {Mingzhen}\ \bibnamefont
  {Tian}},\ }\bibfield  {title} {\enquote {\bibinfo {title} {Robustness of
  single-qubit geometric gate against systematic error},}\ }\href {\doibase
  10.1103/PhysRevA.84.042335} {\bibfield  {journal} {\bibinfo  {journal} {Phys.
  Rev. A}\ }\textbf {\bibinfo {volume} {84}},\ \bibinfo {pages} {042335}
  (\bibinfo {year} {2011})}\BibitemShut {NoStop}%
\bibitem [{\citenamefont {Zhao}\ \emph
  {et~al.}(2017{\natexlab{a}})\citenamefont {Zhao}, \citenamefont {Cui},
  \citenamefont {Xu}, \citenamefont {Sj\"oqvist},\ and\ \citenamefont
  {Tong}}]{Zhao2017}%
  \BibitemOpen
  \bibfield  {author} {\bibinfo {author} {\bibfnamefont {P.~Z.}\ \bibnamefont
  {Zhao}}, \bibinfo {author} {\bibfnamefont {Xiao-Dan}\ \bibnamefont {Cui}},
  \bibinfo {author} {\bibfnamefont {G.~F.}\ \bibnamefont {Xu}}, \bibinfo
  {author} {\bibfnamefont {Erik}\ \bibnamefont {Sj\"oqvist}}, \ and\ \bibinfo
  {author} {\bibfnamefont {D.~M.}\ \bibnamefont {Tong}},\ }\bibfield  {title}
  {\enquote {\bibinfo {title} {Rydberg-atom-based scheme of nonadiabatic
  geometric quantum computation},}\ }\href {\doibase
  10.1103/PhysRevA.96.052316} {\bibfield  {journal} {\bibinfo  {journal} {Phys.
  Rev. A}\ }\textbf {\bibinfo {volume} {96}},\ \bibinfo {pages} {052316}
  (\bibinfo {year} {2017}{\natexlab{a}})}\BibitemShut {NoStop}%
\bibitem [{\citenamefont {Li}\ \emph {et~al.}(2020)\citenamefont {Li},
  \citenamefont {Zhao},\ and\ \citenamefont {Tong}}]{Li2020}%
  \BibitemOpen
  \bibfield  {author} {\bibinfo {author} {\bibfnamefont {K.~Z.}\ \bibnamefont
  {Li}}, \bibinfo {author} {\bibfnamefont {P.~Z.}\ \bibnamefont {Zhao}}, \ and\
  \bibinfo {author} {\bibfnamefont {D.~M.}\ \bibnamefont {Tong}},\ }\bibfield
  {title} {\enquote {\bibinfo {title} {Approach to realizing nonadiabatic
  geometric gates with prescribed evolution paths},}\ }\href {\doibase
  10.1103/PhysRevResearch.2.023295} {\bibfield  {journal} {\bibinfo  {journal}
  {Phys. Rev. Research}\ }\textbf {\bibinfo {volume} {2}},\ \bibinfo {pages}
  {023295} (\bibinfo {year} {2020})}\BibitemShut {NoStop}%
\bibitem [{\citenamefont {Chen}\ and\ \citenamefont {Xue}(2018)}]{Chen2018}%
  \BibitemOpen
  \bibfield  {author} {\bibinfo {author} {\bibfnamefont {Tao}\ \bibnamefont
  {Chen}}\ and\ \bibinfo {author} {\bibfnamefont {Zheng-Yuan}\ \bibnamefont
  {Xue}},\ }\bibfield  {title} {\enquote {\bibinfo {title} {Nonadiabatic
  geometric quantum computation with parametrically tunable coupling},}\ }\href
  {\doibase 10.1103/PhysRevApplied.10.054051} {\bibfield  {journal} {\bibinfo
  {journal} {Phys. Rev. Applied}\ }\textbf {\bibinfo {volume} {10}},\ \bibinfo
  {pages} {054051} (\bibinfo {year} {2018})}\BibitemShut {NoStop}%
\bibitem [{\citenamefont {Zhang}\ \emph {et~al.}(2020)\citenamefont {Zhang},
  \citenamefont {Chen}, \citenamefont {Li}, \citenamefont {Wang},\ and\
  \citenamefont {Xue}}]{Zhang2020}%
  \BibitemOpen
  \bibfield  {author} {\bibinfo {author} {\bibfnamefont {Chengxian}\
  \bibnamefont {Zhang}}, \bibinfo {author} {\bibfnamefont {Tao}\ \bibnamefont
  {Chen}}, \bibinfo {author} {\bibfnamefont {Sai}\ \bibnamefont {Li}}, \bibinfo
  {author} {\bibfnamefont {Xin}\ \bibnamefont {Wang}}, \ and\ \bibinfo {author}
  {\bibfnamefont {Zheng-Yuan}\ \bibnamefont {Xue}},\ }\bibfield  {title}
  {\enquote {\bibinfo {title} {High-fidelity geometric gate for silicon-based
  spin qubits},}\ }\href {\doibase 10.1103/PhysRevA.101.052302} {\bibfield
  {journal} {\bibinfo  {journal} {Phys. Rev. A}\ }\textbf {\bibinfo {volume}
  {101}},\ \bibinfo {pages} {052302} (\bibinfo {year} {2020})}\BibitemShut
  {NoStop}%
\bibitem [{\citenamefont {Liu}\ \emph {et~al.}(2019)\citenamefont {Liu},
  \citenamefont {Song}, \citenamefont {Xue}, \citenamefont {Wang},\ and\
  \citenamefont {Yung}}]{Liu2019}%
  \BibitemOpen
  \bibfield  {author} {\bibinfo {author} {\bibfnamefont {Bao-Jie}\ \bibnamefont
  {Liu}}, \bibinfo {author} {\bibfnamefont {Xue-Ke}\ \bibnamefont {Song}},
  \bibinfo {author} {\bibfnamefont {Zheng-Yuan}\ \bibnamefont {Xue}}, \bibinfo
  {author} {\bibfnamefont {Xin}\ \bibnamefont {Wang}}, \ and\ \bibinfo {author}
  {\bibfnamefont {Man-Hong}\ \bibnamefont {Yung}},\ }\bibfield  {title}
  {\enquote {\bibinfo {title} {Plug-and-play approach to nonadiabatic geometric
  quantum gates},}\ }\href {\doibase 10.1103/PhysRevLett.123.100501} {\bibfield
   {journal} {\bibinfo  {journal} {Phys. Rev. Lett.}\ }\textbf {\bibinfo
  {volume} {123}},\ \bibinfo {pages} {100501} (\bibinfo {year}
  {2019})}\BibitemShut {NoStop}%
\bibitem [{\citenamefont {Sjöqvist}\ \emph {et~al.}(2012)\citenamefont
  {Sjöqvist}, \citenamefont {Tong}, \citenamefont {Mauritz~Andersson},
  \citenamefont {Hessmo}, \citenamefont {Johansson},\ and\ \citenamefont
  {Singh}}]{Sjoeqvist2012}%
  \BibitemOpen
  \bibfield  {author} {\bibinfo {author} {\bibfnamefont {Erik}\ \bibnamefont
  {Sjöqvist}}, \bibinfo {author} {\bibfnamefont {D.~M.}\ \bibnamefont {Tong}},
  \bibinfo {author} {\bibfnamefont {L.}~\bibnamefont {Mauritz~Andersson}},
  \bibinfo {author} {\bibfnamefont {Björn}\ \bibnamefont {Hessmo}}, \bibinfo
  {author} {\bibfnamefont {Markus}\ \bibnamefont {Johansson}}, \ and\ \bibinfo
  {author} {\bibfnamefont {Kuldip}\ \bibnamefont {Singh}},\ }\bibfield  {title}
  {\enquote {\bibinfo {title} {Non-adiabatic holonomic quantum computation},}\
  }\href {http://dx.doi.org/10.1088/1367-2630/14/10/103035} {\bibfield
  {journal} {\bibinfo  {journal} {New Journal of Physics}\ }\textbf {\bibinfo
  {volume} {14}},\ \bibinfo {pages} {103035} (\bibinfo {year}
  {2012})}\BibitemShut {NoStop}%
\bibitem [{\citenamefont {Xu}\ \emph {et~al.}(2012{\natexlab{b}})\citenamefont
  {Xu}, \citenamefont {Zhang}, \citenamefont {Tong}, \citenamefont
  {Sj\"oqvist},\ and\ \citenamefont {Kwek}}]{Xugf2012}%
  \BibitemOpen
  \bibfield  {author} {\bibinfo {author} {\bibfnamefont {G.~F.}\ \bibnamefont
  {Xu}}, \bibinfo {author} {\bibfnamefont {J.}~\bibnamefont {Zhang}}, \bibinfo
  {author} {\bibfnamefont {D.~M.}\ \bibnamefont {Tong}}, \bibinfo {author}
  {\bibfnamefont {Erik}\ \bibnamefont {Sj\"oqvist}}, \ and\ \bibinfo {author}
  {\bibfnamefont {L.~C.}\ \bibnamefont {Kwek}},\ }\bibfield  {title} {\enquote
  {\bibinfo {title} {Nonadiabatic holonomic quantum computation in
  decoherence-free subspaces},}\ }\href {\doibase
  10.1103/PhysRevLett.109.170501} {\bibfield  {journal} {\bibinfo  {journal}
  {Phys. Rev. Lett.}\ }\textbf {\bibinfo {volume} {109}},\ \bibinfo {pages}
  {170501} (\bibinfo {year} {2012}{\natexlab{b}})}\BibitemShut {NoStop}%
\bibitem [{\citenamefont {Xue}\ \emph {et~al.}(2015)\citenamefont {Xue},
  \citenamefont {Zhou},\ and\ \citenamefont {Wang}}]{Xue2015}%
  \BibitemOpen
  \bibfield  {author} {\bibinfo {author} {\bibfnamefont {Zheng-Yuan}\
  \bibnamefont {Xue}}, \bibinfo {author} {\bibfnamefont {Jian}\ \bibnamefont
  {Zhou}}, \ and\ \bibinfo {author} {\bibfnamefont {Z.~D.}\ \bibnamefont
  {Wang}},\ }\bibfield  {title} {\enquote {\bibinfo {title} {Universal
  holonomic quantum gates in decoherence-free subspace on superconducting
  circuits},}\ }\href {\doibase 10.1103/PhysRevA.92.022320} {\bibfield
  {journal} {\bibinfo  {journal} {Phys. Rev. A}\ }\textbf {\bibinfo {volume}
  {92}},\ \bibinfo {pages} {022320} (\bibinfo {year} {2015})}\BibitemShut
  {NoStop}%
\bibitem [{\citenamefont {Xue}\ \emph {et~al.}(2017)\citenamefont {Xue},
  \citenamefont {Gu}, \citenamefont {Hong}, \citenamefont {Yang}, \citenamefont
  {Zhang}, \citenamefont {Hu},\ and\ \citenamefont {You}}]{Xue2017}%
  \BibitemOpen
  \bibfield  {author} {\bibinfo {author} {\bibfnamefont {Zheng-Yuan}\
  \bibnamefont {Xue}}, \bibinfo {author} {\bibfnamefont {Feng-Lei}\
  \bibnamefont {Gu}}, \bibinfo {author} {\bibfnamefont {Zhuo-Ping}\
  \bibnamefont {Hong}}, \bibinfo {author} {\bibfnamefont {Zi-He}\ \bibnamefont
  {Yang}}, \bibinfo {author} {\bibfnamefont {Dan-Wei}\ \bibnamefont {Zhang}},
  \bibinfo {author} {\bibfnamefont {Yong}\ \bibnamefont {Hu}}, \ and\ \bibinfo
  {author} {\bibfnamefont {J.~Q.}\ \bibnamefont {You}},\ }\bibfield  {title}
  {\enquote {\bibinfo {title} {Nonadiabatic holonomic quantum computation with
  dressed-state qubits},}\ }\href {\doibase 10.1103/PhysRevApplied.7.054022}
  {\bibfield  {journal} {\bibinfo  {journal} {Phys. Rev. Applied}\ }\textbf
  {\bibinfo {volume} {7}},\ \bibinfo {pages} {054022} (\bibinfo {year}
  {2017})}\BibitemShut {NoStop}%
\bibitem [{\citenamefont {Zhou}\ \emph
  {et~al.}(2017{\natexlab{a}})\citenamefont {Zhou}, \citenamefont {Liu},
  \citenamefont {Hong},\ and\ \citenamefont {Xue}}]{Zhou2017}%
  \BibitemOpen
  \bibfield  {author} {\bibinfo {author} {\bibfnamefont {Jian}\ \bibnamefont
  {Zhou}}, \bibinfo {author} {\bibfnamefont {BaoJie}\ \bibnamefont {Liu}},
  \bibinfo {author} {\bibfnamefont {ZhuoPing}\ \bibnamefont {Hong}}, \ and\
  \bibinfo {author} {\bibfnamefont {ZhengYuan}\ \bibnamefont {Xue}},\
  }\bibfield  {title} {\enquote {\bibinfo {title} {Fast holonomic quantum
  computation based on solid-state spins with all-optical control},}\ }\href
  {https://doi.org/10.1007/s11433-017-9119-8} {\bibfield  {journal} {\bibinfo
  {journal} {Science China Physics, Mechanics \& Astronomy}\ }\textbf {\bibinfo
  {volume} {61}},\ \bibinfo {pages} {010312} (\bibinfo {year}
  {2017}{\natexlab{a}})}\BibitemShut {NoStop}%
\bibitem [{\citenamefont {Hong}\ \emph {et~al.}(2018)\citenamefont {Hong},
  \citenamefont {Liu}, \citenamefont {Cai}, \citenamefont {Zhang},
  \citenamefont {Hu}, \citenamefont {Wang},\ and\ \citenamefont
  {Xue}}]{Hong2018}%
  \BibitemOpen
  \bibfield  {author} {\bibinfo {author} {\bibfnamefont {Zhuo-Ping}\
  \bibnamefont {Hong}}, \bibinfo {author} {\bibfnamefont {Bao-Jie}\
  \bibnamefont {Liu}}, \bibinfo {author} {\bibfnamefont {Jia-Qi}\ \bibnamefont
  {Cai}}, \bibinfo {author} {\bibfnamefont {Xin-Ding}\ \bibnamefont {Zhang}},
  \bibinfo {author} {\bibfnamefont {Yong}\ \bibnamefont {Hu}}, \bibinfo
  {author} {\bibfnamefont {Z.~D.}\ \bibnamefont {Wang}}, \ and\ \bibinfo
  {author} {\bibfnamefont {Zheng-Yuan}\ \bibnamefont {Xue}},\ }\bibfield
  {title} {\enquote {\bibinfo {title} {Implementing universal nonadiabatic
  holonomic quantum gates with transmons},}\ }\href {\doibase
  10.1103/PhysRevA.97.022332} {\bibfield  {journal} {\bibinfo  {journal} {Phys.
  Rev. A}\ }\textbf {\bibinfo {volume} {97}},\ \bibinfo {pages} {022332}
  (\bibinfo {year} {2018})}\BibitemShut {NoStop}%
\bibitem [{\citenamefont {Azimi~Mousolou}(2017)}]{AzimiMousolou2017}%
  \BibitemOpen
  \bibfield  {author} {\bibinfo {author} {\bibfnamefont {Vahid}\ \bibnamefont
  {Azimi~Mousolou}},\ }\bibfield  {title} {\enquote {\bibinfo {title} {Electric
  nonadiabatic geometric entangling gates on spin qubits},}\ }\href {\doibase
  10.1103/PhysRevA.96.012307} {\bibfield  {journal} {\bibinfo  {journal} {Phys.
  Rev. A}\ }\textbf {\bibinfo {volume} {96}},\ \bibinfo {pages} {012307}
  (\bibinfo {year} {2017})}\BibitemShut {NoStop}%
\bibitem [{\citenamefont {Zhao}\ \emph {et~al.}(2020)\citenamefont {Zhao},
  \citenamefont {Li}, \citenamefont {Xu},\ and\ \citenamefont
  {Tong}}]{Zhao2020}%
  \BibitemOpen
  \bibfield  {author} {\bibinfo {author} {\bibfnamefont {P.~Z.}\ \bibnamefont
  {Zhao}}, \bibinfo {author} {\bibfnamefont {K.~Z.}\ \bibnamefont {Li}},
  \bibinfo {author} {\bibfnamefont {G.~F.}\ \bibnamefont {Xu}}, \ and\ \bibinfo
  {author} {\bibfnamefont {D.~M.}\ \bibnamefont {Tong}},\ }\bibfield  {title}
  {\enquote {\bibinfo {title} {General approach for constructing hamiltonians
  for nonadiabatic holonomic quantum computation},}\ }\href {\doibase
  10.1103/PhysRevA.101.062306} {\bibfield  {journal} {\bibinfo  {journal}
  {Phys. Rev. A}\ }\textbf {\bibinfo {volume} {101}},\ \bibinfo {pages}
  {062306} (\bibinfo {year} {2020})}\BibitemShut {NoStop}%
\bibitem [{\citenamefont {Johansson}\ \emph {et~al.}(2012)\citenamefont
  {Johansson}, \citenamefont {Sj\"oqvist}, \citenamefont {Andersson},
  \citenamefont {Ericsson}, \citenamefont {Hessmo}, \citenamefont {Singh},\
  and\ \citenamefont {Tong}}]{Johansson2012}%
  \BibitemOpen
  \bibfield  {author} {\bibinfo {author} {\bibfnamefont {Markus}\ \bibnamefont
  {Johansson}}, \bibinfo {author} {\bibfnamefont {Erik}\ \bibnamefont
  {Sj\"oqvist}}, \bibinfo {author} {\bibfnamefont {L.~Mauritz}\ \bibnamefont
  {Andersson}}, \bibinfo {author} {\bibfnamefont {Marie}\ \bibnamefont
  {Ericsson}}, \bibinfo {author} {\bibfnamefont {Bj\"orn}\ \bibnamefont
  {Hessmo}}, \bibinfo {author} {\bibfnamefont {Kuldip}\ \bibnamefont {Singh}},
  \ and\ \bibinfo {author} {\bibfnamefont {D.~M.}\ \bibnamefont {Tong}},\
  }\bibfield  {title} {\enquote {\bibinfo {title} {Robustness of nonadiabatic
  holonomic gates},}\ }\href {\doibase 10.1103/PhysRevA.86.062322} {\bibfield
  {journal} {\bibinfo  {journal} {Phys. Rev. A}\ }\textbf {\bibinfo {volume}
  {86}},\ \bibinfo {pages} {062322} (\bibinfo {year} {2012})}\BibitemShut
  {NoStop}%
\bibitem [{\citenamefont {Zheng}\ \emph {et~al.}(2016)\citenamefont {Zheng},
  \citenamefont {Yang},\ and\ \citenamefont {Nori}}]{Zheng2016}%
  \BibitemOpen
  \bibfield  {author} {\bibinfo {author} {\bibfnamefont {Shi-Biao}\
  \bibnamefont {Zheng}}, \bibinfo {author} {\bibfnamefont {Chui-Ping}\
  \bibnamefont {Yang}}, \ and\ \bibinfo {author} {\bibfnamefont {Franco}\
  \bibnamefont {Nori}},\ }\bibfield  {title} {\enquote {\bibinfo {title}
  {Comparison of the sensitivity to systematic errors between nonadiabatic
  non-{Abelian} geometric gates and their dynamical counterparts},}\ }\href
  {\doibase 10.1103/PhysRevA.93.032313} {\bibfield  {journal} {\bibinfo
  {journal} {Phys. Rev. A}\ }\textbf {\bibinfo {volume} {93}},\ \bibinfo
  {pages} {032313} (\bibinfo {year} {2016})}\BibitemShut {NoStop}%
\bibitem [{\citenamefont {Ramberg}\ and\ \citenamefont
  {Sj\"oqvist}(2019)}]{Ramberg2019}%
  \BibitemOpen
  \bibfield  {author} {\bibinfo {author} {\bibfnamefont {Nicklas}\ \bibnamefont
  {Ramberg}}\ and\ \bibinfo {author} {\bibfnamefont {Erik}\ \bibnamefont
  {Sj\"oqvist}},\ }\bibfield  {title} {\enquote {\bibinfo {title}
  {Environment-assisted holonomic quantum maps},}\ }\href {\doibase
  10.1103/PhysRevLett.122.140501} {\bibfield  {journal} {\bibinfo  {journal}
  {Phys. Rev. Lett.}\ }\textbf {\bibinfo {volume} {122}},\ \bibinfo {pages}
  {140501} (\bibinfo {year} {2019})}\BibitemShut {NoStop}%
\bibitem [{\citenamefont {Jing}\ \emph {et~al.}(2017)\citenamefont {Jing},
  \citenamefont {Lam},\ and\ \citenamefont {Wu}}]{Jing2017}%
  \BibitemOpen
  \bibfield  {author} {\bibinfo {author} {\bibfnamefont {Jun}\ \bibnamefont
  {Jing}}, \bibinfo {author} {\bibfnamefont {Chi-Hang}\ \bibnamefont {Lam}}, \
  and\ \bibinfo {author} {\bibfnamefont {Lian-Ao}\ \bibnamefont {Wu}},\
  }\bibfield  {title} {\enquote {\bibinfo {title} {Non-{Abelian} holonomic
  transformation in the presence of classical noise},}\ }\href {\doibase
  10.1103/PhysRevA.95.012334} {\bibfield  {journal} {\bibinfo  {journal} {Phys.
  Rev. A}\ }\textbf {\bibinfo {volume} {95}},\ \bibinfo {pages} {012334}
  (\bibinfo {year} {2017})}\BibitemShut {NoStop}%
\bibitem [{\citenamefont {Liu}\ \emph {et~al.}(2020{\natexlab{a}})\citenamefont
  {Liu}, \citenamefont {Wang},\ and\ \citenamefont {Yung}}]{Liu2020}%
  \BibitemOpen
  \bibfield  {author} {\bibinfo {author} {\bibfnamefont {Bao-Jie}\ \bibnamefont
  {Liu}}, \bibinfo {author} {\bibfnamefont {Yuan-Sheng}\ \bibnamefont {Wang}},
  \ and\ \bibinfo {author} {\bibfnamefont {Man-Hong}\ \bibnamefont {Yung}},\
  }\href@noop {} {\enquote {\bibinfo {title} {Global property condition-based
  non-adiabatic geometric quantum control},}\ } (\bibinfo {year}
  {2020}{\natexlab{a}}),\ \Eprint {http://arxiv.org/abs/2008.02176}
  {arXiv:2008.02176 [quant-ph]} \BibitemShut {NoStop}%
\bibitem [{\citenamefont {Feng}\ \emph {et~al.}(2013)\citenamefont {Feng},
  \citenamefont {Xu},\ and\ \citenamefont {Long}}]{Feng2013}%
  \BibitemOpen
  \bibfield  {author} {\bibinfo {author} {\bibfnamefont {Guanru}\ \bibnamefont
  {Feng}}, \bibinfo {author} {\bibfnamefont {Guofu}\ \bibnamefont {Xu}}, \ and\
  \bibinfo {author} {\bibfnamefont {Guilu}\ \bibnamefont {Long}},\ }\bibfield
  {title} {\enquote {\bibinfo {title} {Experimental realization of nonadiabatic
  holonomic quantum computation},}\ }\href {\doibase
  10.1103/PhysRevLett.110.190501} {\bibfield  {journal} {\bibinfo  {journal}
  {Phys. Rev. Lett.}\ }\textbf {\bibinfo {volume} {110}},\ \bibinfo {pages}
  {190501} (\bibinfo {year} {2013})}\BibitemShut {NoStop}%
\bibitem [{\citenamefont {Li}\ \emph {et~al.}(2017)\citenamefont {Li},
  \citenamefont {Liu},\ and\ \citenamefont {Long}}]{Li2017}%
  \BibitemOpen
  \bibfield  {author} {\bibinfo {author} {\bibfnamefont {Hang}\ \bibnamefont
  {Li}}, \bibinfo {author} {\bibfnamefont {Yang}\ \bibnamefont {Liu}}, \ and\
  \bibinfo {author} {\bibfnamefont {GuiLu}\ \bibnamefont {Long}},\ }\bibfield
  {title} {\enquote {\bibinfo {title} {Experimental realization of single-shot
  nonadiabatic holonomic gates in nuclear spins},}\ }\href
  {https://doi.org/10.1007/s11433-017-9058-7} {\bibfield  {journal} {\bibinfo
  {journal} {Science China Physics, Mechanics \& Astronomy}\ }\textbf {\bibinfo
  {volume} {60}},\ \bibinfo {pages} {080311} (\bibinfo {year}
  {2017})}\BibitemShut {NoStop}%
\bibitem [{\citenamefont {Abdumalikov~Jr}\ \emph {et~al.}(2013)\citenamefont
  {Abdumalikov~Jr}, \citenamefont {Fink}, \citenamefont {Juliusson},
  \citenamefont {Pechal}, \citenamefont {Berger}, \citenamefont {Wallraff},\
  and\ \citenamefont {Filipp}}]{AbdumalikovJr2013}%
  \BibitemOpen
  \bibfield  {author} {\bibinfo {author} {\bibfnamefont {A.~A.}\ \bibnamefont
  {Abdumalikov~Jr}}, \bibinfo {author} {\bibfnamefont {J.~M.}\ \bibnamefont
  {Fink}}, \bibinfo {author} {\bibfnamefont {K.}~\bibnamefont {Juliusson}},
  \bibinfo {author} {\bibfnamefont {M.}~\bibnamefont {Pechal}}, \bibinfo
  {author} {\bibfnamefont {S.}~\bibnamefont {Berger}}, \bibinfo {author}
  {\bibfnamefont {A.}~\bibnamefont {Wallraff}}, \ and\ \bibinfo {author}
  {\bibfnamefont {S.}~\bibnamefont {Filipp}},\ }\bibfield  {title} {\enquote
  {\bibinfo {title} {Experimental realization of non-{Abelian} non-adiabatic
  geometric gates},}\ }\href {https://doi.org/10.1038/nature12010} {\bibfield
  {journal} {\bibinfo  {journal} {Nature}\ }\textbf {\bibinfo {volume} {496}},\
  \bibinfo {pages} {482--485} (\bibinfo {year} {2013})}\BibitemShut {NoStop}%
\bibitem [{\citenamefont {Xu}\ \emph {et~al.}(2018)\citenamefont {Xu},
  \citenamefont {Cai}, \citenamefont {Ma}, \citenamefont {Mu}, \citenamefont
  {Hu}, \citenamefont {Chen}, \citenamefont {Wang}, \citenamefont {Song},
  \citenamefont {Xue}, \citenamefont {Yin},\ and\ \citenamefont
  {Sun}}]{Xu2018}%
  \BibitemOpen
  \bibfield  {author} {\bibinfo {author} {\bibfnamefont {Y.}~\bibnamefont
  {Xu}}, \bibinfo {author} {\bibfnamefont {W.}~\bibnamefont {Cai}}, \bibinfo
  {author} {\bibfnamefont {Y.}~\bibnamefont {Ma}}, \bibinfo {author}
  {\bibfnamefont {X.}~\bibnamefont {Mu}}, \bibinfo {author} {\bibfnamefont
  {L.}~\bibnamefont {Hu}}, \bibinfo {author} {\bibfnamefont {Tao}\ \bibnamefont
  {Chen}}, \bibinfo {author} {\bibfnamefont {H.}~\bibnamefont {Wang}}, \bibinfo
  {author} {\bibfnamefont {Y.~P.}\ \bibnamefont {Song}}, \bibinfo {author}
  {\bibfnamefont {Zheng-Yuan}\ \bibnamefont {Xue}}, \bibinfo {author}
  {\bibfnamefont {Zhang-qi}\ \bibnamefont {Yin}}, \ and\ \bibinfo {author}
  {\bibfnamefont {L.}~\bibnamefont {Sun}},\ }\bibfield  {title} {\enquote
  {\bibinfo {title} {Single-loop realization of arbitrary nonadiabatic
  holonomic single-qubit quantum gates in a superconducting circuit},}\ }\href
  {\doibase 10.1103/PhysRevLett.121.110501} {\bibfield  {journal} {\bibinfo
  {journal} {Phys. Rev. Lett.}\ }\textbf {\bibinfo {volume} {121}},\ \bibinfo
  {pages} {110501} (\bibinfo {year} {2018})}\BibitemShut {NoStop}%
\bibitem [{\citenamefont {Zhang}\ \emph {et~al.}(2019)\citenamefont {Zhang},
  \citenamefont {Zhao}, \citenamefont {Wang}, \citenamefont {Xiang},
  \citenamefont {Jia}, \citenamefont {Duan}, \citenamefont {Tong},
  \citenamefont {Yin},\ and\ \citenamefont {Guo}}]{Zhang2019}%
  \BibitemOpen
  \bibfield  {author} {\bibinfo {author} {\bibfnamefont {Zhenxing}\
  \bibnamefont {Zhang}}, \bibinfo {author} {\bibfnamefont {P.~Z.}\ \bibnamefont
  {Zhao}}, \bibinfo {author} {\bibfnamefont {Tenghui}\ \bibnamefont {Wang}},
  \bibinfo {author} {\bibfnamefont {Liang}\ \bibnamefont {Xiang}}, \bibinfo
  {author} {\bibfnamefont {Zhilong}\ \bibnamefont {Jia}}, \bibinfo {author}
  {\bibfnamefont {Peng}\ \bibnamefont {Duan}}, \bibinfo {author} {\bibfnamefont
  {D.~M.}\ \bibnamefont {Tong}}, \bibinfo {author} {\bibfnamefont
  {Yi}~\bibnamefont {Yin}}, \ and\ \bibinfo {author} {\bibfnamefont {Guoping}\
  \bibnamefont {Guo}},\ }\bibfield  {title} {\enquote {\bibinfo {title}
  {Single-shot realization of nonadiabatic holonomic gates with a
  superconducting xmon qutrit},}\ }\href
  {http://dx.doi.org/10.1088/1367-2630/ab2e26} {\bibfield  {journal} {\bibinfo
  {journal} {New Journal of Physics}\ }\textbf {\bibinfo {volume} {21}},\
  \bibinfo {pages} {073024} (\bibinfo {year} {2019})}\BibitemShut {NoStop}%
\bibitem [{\citenamefont {Danilin}\ \emph {et~al.}(2018)\citenamefont
  {Danilin}, \citenamefont {Vepsäläinen},\ and\ \citenamefont
  {Paraoanu}}]{Danilin2018}%
  \BibitemOpen
  \bibfield  {author} {\bibinfo {author} {\bibfnamefont {S.}~\bibnamefont
  {Danilin}}, \bibinfo {author} {\bibfnamefont {A.}~\bibnamefont
  {Vepsäläinen}}, \ and\ \bibinfo {author} {\bibfnamefont {G.~S.}\
  \bibnamefont {Paraoanu}},\ }\bibfield  {title} {\enquote {\bibinfo {title}
  {Experimental state control by fast non-abelian holonomic gates with a
  superconducting qutrit},}\ }\href
  {http://dx.doi.org/10.1088/1402-4896/aab084} {\bibfield  {journal} {\bibinfo
  {journal} {Physica Scripta}\ }\textbf {\bibinfo {volume} {93}},\ \bibinfo
  {pages} {055101} (\bibinfo {year} {2018})}\BibitemShut {NoStop}%
\bibitem [{\citenamefont {Egger}\ \emph {et~al.}(2019)\citenamefont {Egger},
  \citenamefont {Ganzhorn}, \citenamefont {Salis}, \citenamefont {Fuhrer},
  \citenamefont {M\"uller}, \citenamefont {Barkoutsos}, \citenamefont {Moll},
  \citenamefont {Tavernelli},\ and\ \citenamefont {Filipp}}]{Egger2019}%
  \BibitemOpen
  \bibfield  {author} {\bibinfo {author} {\bibfnamefont {D.J.}\ \bibnamefont
  {Egger}}, \bibinfo {author} {\bibfnamefont {M.}~\bibnamefont {Ganzhorn}},
  \bibinfo {author} {\bibfnamefont {G.}~\bibnamefont {Salis}}, \bibinfo
  {author} {\bibfnamefont {A.}~\bibnamefont {Fuhrer}}, \bibinfo {author}
  {\bibfnamefont {P.}~\bibnamefont {M\"uller}}, \bibinfo {author}
  {\bibfnamefont {P.Kl.}\ \bibnamefont {Barkoutsos}}, \bibinfo {author}
  {\bibfnamefont {N.}~\bibnamefont {Moll}}, \bibinfo {author} {\bibfnamefont
  {I.}~\bibnamefont {Tavernelli}}, \ and\ \bibinfo {author} {\bibfnamefont
  {S.}~\bibnamefont {Filipp}},\ }\bibfield  {title} {\enquote {\bibinfo {title}
  {Entanglement generation in superconducting qubits using holonomic
  operations},}\ }\href {\doibase 10.1103/PhysRevApplied.11.014017} {\bibfield
  {journal} {\bibinfo  {journal} {Phys. Rev. Applied}\ }\textbf {\bibinfo
  {volume} {11}},\ \bibinfo {pages} {014017} (\bibinfo {year}
  {2019})}\BibitemShut {NoStop}%
\bibitem [{\citenamefont {Zu}\ \emph {et~al.}(2014)\citenamefont {Zu},
  \citenamefont {Wang}, \citenamefont {He}, \citenamefont {Zhang},
  \citenamefont {Dai}, \citenamefont {Wang},\ and\ \citenamefont
  {Duan}}]{Zu2014}%
  \BibitemOpen
  \bibfield  {author} {\bibinfo {author} {\bibfnamefont {C.}~\bibnamefont
  {Zu}}, \bibinfo {author} {\bibfnamefont {W.-B.}\ \bibnamefont {Wang}},
  \bibinfo {author} {\bibfnamefont {L.}~\bibnamefont {He}}, \bibinfo {author}
  {\bibfnamefont {W.-G.}\ \bibnamefont {Zhang}}, \bibinfo {author}
  {\bibfnamefont {C.-Y.}\ \bibnamefont {Dai}}, \bibinfo {author} {\bibfnamefont
  {F.}~\bibnamefont {Wang}}, \ and\ \bibinfo {author} {\bibfnamefont {L.-M.}\
  \bibnamefont {Duan}},\ }\bibfield  {title} {\enquote {\bibinfo {title}
  {Experimental realization of universal geometric quantum gates with
  solid-state spins},}\ }\href {https://doi.org/10.1038/nature13729} {\bibfield
   {journal} {\bibinfo  {journal} {Nature}\ }\textbf {\bibinfo {volume}
  {514}},\ \bibinfo {pages} {72--75} (\bibinfo {year} {2014})}\BibitemShut
  {NoStop}%
\bibitem [{\citenamefont {Arroyo-Camejo}\ \emph {et~al.}(2014)\citenamefont
  {Arroyo-Camejo}, \citenamefont {Lazariev}, \citenamefont {Hell},\ and\
  \citenamefont {Balasubramanian}}]{Arroyo-Camejo2014}%
  \BibitemOpen
  \bibfield  {author} {\bibinfo {author} {\bibfnamefont {Silvia}\ \bibnamefont
  {Arroyo-Camejo}}, \bibinfo {author} {\bibfnamefont {Andrii}\ \bibnamefont
  {Lazariev}}, \bibinfo {author} {\bibfnamefont {Stefan~W.}\ \bibnamefont
  {Hell}}, \ and\ \bibinfo {author} {\bibfnamefont {Gopalakrishnan}\
  \bibnamefont {Balasubramanian}},\ }\bibfield  {title} {\enquote {\bibinfo
  {title} {Room temperature high-fidelity holonomic single-qubit gate on a
  solid-state spin},}\ }\href {https://doi.org/10.1038/ncomms5870} {\bibfield
  {journal} {\bibinfo  {journal} {Nature Communications}\ }\textbf {\bibinfo
  {volume} {5}},\ \bibinfo {pages} {4870} (\bibinfo {year} {2014})}\BibitemShut
  {NoStop}%
\bibitem [{\citenamefont {Zhou}\ \emph
  {et~al.}(2017{\natexlab{b}})\citenamefont {Zhou}, \citenamefont {Jerger},
  \citenamefont {Shkolnikov}, \citenamefont {Heremans}, \citenamefont
  {Burkard},\ and\ \citenamefont {Awschalom}}]{Zhou2017a}%
  \BibitemOpen
  \bibfield  {author} {\bibinfo {author} {\bibfnamefont {Brian~B.}\
  \bibnamefont {Zhou}}, \bibinfo {author} {\bibfnamefont {Paul~C.}\
  \bibnamefont {Jerger}}, \bibinfo {author} {\bibfnamefont {V.~O.}\
  \bibnamefont {Shkolnikov}}, \bibinfo {author} {\bibfnamefont {F.~Joseph}\
  \bibnamefont {Heremans}}, \bibinfo {author} {\bibfnamefont {Guido}\
  \bibnamefont {Burkard}}, \ and\ \bibinfo {author} {\bibfnamefont {David~D.}\
  \bibnamefont {Awschalom}},\ }\bibfield  {title} {\enquote {\bibinfo {title}
  {Holonomic quantum control by coherent optical excitation in diamond},}\
  }\href {\doibase 10.1103/PhysRevLett.119.140503} {\bibfield  {journal}
  {\bibinfo  {journal} {Phys. Rev. Lett.}\ }\textbf {\bibinfo {volume} {119}},\
  \bibinfo {pages} {140503} (\bibinfo {year} {2017}{\natexlab{b}})}\BibitemShut
  {NoStop}%
\bibitem [{\citenamefont {Sekiguchi}\ \emph {et~al.}(2017)\citenamefont
  {Sekiguchi}, \citenamefont {Niikura}, \citenamefont {Kuroiwa}, \citenamefont
  {Kano},\ and\ \citenamefont {Kosaka}}]{Sekiguchi2017}%
  \BibitemOpen
  \bibfield  {author} {\bibinfo {author} {\bibfnamefont {Yuhei}\ \bibnamefont
  {Sekiguchi}}, \bibinfo {author} {\bibfnamefont {Naeko}\ \bibnamefont
  {Niikura}}, \bibinfo {author} {\bibfnamefont {Ryota}\ \bibnamefont
  {Kuroiwa}}, \bibinfo {author} {\bibfnamefont {Hiroki}\ \bibnamefont {Kano}},
  \ and\ \bibinfo {author} {\bibfnamefont {Hideo}\ \bibnamefont {Kosaka}},\
  }\bibfield  {title} {\enquote {\bibinfo {title} {Optical holonomic single
  quantum gates with a geometric spin under a zero field},}\ }\href
  {https://doi.org/10.1038/nphoton.2017.40} {\bibfield  {journal} {\bibinfo
  {journal} {Nature Photonics}\ }\textbf {\bibinfo {volume} {11}},\ \bibinfo
  {pages} {309--314} (\bibinfo {year} {2017})}\BibitemShut {NoStop}%
\bibitem [{\citenamefont {Nagata}\ \emph {et~al.}(2018)\citenamefont {Nagata},
  \citenamefont {Kuramitani}, \citenamefont {Sekiguchi},\ and\ \citenamefont
  {Kosaka}}]{Nagata2018}%
  \BibitemOpen
  \bibfield  {author} {\bibinfo {author} {\bibfnamefont {Kodai}\ \bibnamefont
  {Nagata}}, \bibinfo {author} {\bibfnamefont {Kouyou}\ \bibnamefont
  {Kuramitani}}, \bibinfo {author} {\bibfnamefont {Yuhei}\ \bibnamefont
  {Sekiguchi}}, \ and\ \bibinfo {author} {\bibfnamefont {Hideo}\ \bibnamefont
  {Kosaka}},\ }\bibfield  {title} {\enquote {\bibinfo {title} {Universal
  holonomic quantum gates over geometric spin qubits with polarised
  microwaves},}\ }\href {https://doi.org/10.1038/s41467-018-05664-w} {\bibfield
   {journal} {\bibinfo  {journal} {Nature Communications}\ }\textbf {\bibinfo
  {volume} {9}},\ \bibinfo {pages} {3227} (\bibinfo {year} {2018})}\BibitemShut
  {NoStop}%
\bibitem [{\citenamefont {Ishida}\ \emph {et~al.}(2018)\citenamefont {Ishida},
  \citenamefont {Nakamura}, \citenamefont {Tanaka}, \citenamefont {Mishima},
  \citenamefont {Kano}, \citenamefont {Kuroiwa}, \citenamefont {Sekiguchi},\
  and\ \citenamefont {Kosaka}}]{Ishida:18}%
  \BibitemOpen
  \bibfield  {author} {\bibinfo {author} {\bibfnamefont {Naoki}\ \bibnamefont
  {Ishida}}, \bibinfo {author} {\bibfnamefont {Takaaki}\ \bibnamefont
  {Nakamura}}, \bibinfo {author} {\bibfnamefont {Touta}\ \bibnamefont
  {Tanaka}}, \bibinfo {author} {\bibfnamefont {Shota}\ \bibnamefont {Mishima}},
  \bibinfo {author} {\bibfnamefont {Hiroki}\ \bibnamefont {Kano}}, \bibinfo
  {author} {\bibfnamefont {Ryota}\ \bibnamefont {Kuroiwa}}, \bibinfo {author}
  {\bibfnamefont {Yuhei}\ \bibnamefont {Sekiguchi}}, \ and\ \bibinfo {author}
  {\bibfnamefont {Hideo}\ \bibnamefont {Kosaka}},\ }\bibfield  {title}
  {\enquote {\bibinfo {title} {Universal holonomic single quantum gates over a
  geometric spin with phase-modulated polarized light},}\ }\href {\doibase
  10.1364/OL.43.002380} {\bibfield  {journal} {\bibinfo  {journal} {Opt.
  Lett.}\ }\textbf {\bibinfo {volume} {43}},\ \bibinfo {pages} {2380--2383}
  (\bibinfo {year} {2018})}\BibitemShut {NoStop}%
\bibitem [{\citenamefont {Zhu}\ and\ \citenamefont {Wang}(2003)}]{Zhu2003}%
  \BibitemOpen
  \bibfield  {author} {\bibinfo {author} {\bibfnamefont {Shi-Liang}\
  \bibnamefont {Zhu}}\ and\ \bibinfo {author} {\bibfnamefont {Z.~D.}\
  \bibnamefont {Wang}},\ }\bibfield  {title} {\enquote {\bibinfo {title}
  {Unconventional geometric quantum computation},}\ }\href {\doibase
  10.1103/PhysRevLett.91.187902} {\bibfield  {journal} {\bibinfo  {journal}
  {Phys. Rev. Lett.}\ }\textbf {\bibinfo {volume} {91}},\ \bibinfo {pages}
  {187902} (\bibinfo {year} {2003})}\BibitemShut {NoStop}%
\bibitem [{\citenamefont {Feng}\ \emph {et~al.}(2009)\citenamefont {Feng},
  \citenamefont {Wu}, \citenamefont {Sun},\ and\ \citenamefont
  {Oh}}]{Feng2009}%
  \BibitemOpen
  \bibfield  {author} {\bibinfo {author} {\bibfnamefont {Xun-Li}\ \bibnamefont
  {Feng}}, \bibinfo {author} {\bibfnamefont {Chunfeng}\ \bibnamefont {Wu}},
  \bibinfo {author} {\bibfnamefont {Hui}\ \bibnamefont {Sun}}, \ and\ \bibinfo
  {author} {\bibfnamefont {C.~H.}\ \bibnamefont {Oh}},\ }\bibfield  {title}
  {\enquote {\bibinfo {title} {Geometric entangling gates in decoherence-free
  subspaces with minimal requirements},}\ }\href {\doibase
  10.1103/PhysRevLett.103.200501} {\bibfield  {journal} {\bibinfo  {journal}
  {Phys. Rev. Lett.}\ }\textbf {\bibinfo {volume} {103}},\ \bibinfo {pages}
  {200501} (\bibinfo {year} {2009})}\BibitemShut {NoStop}%
\bibitem [{\citenamefont {Ota}\ and\ \citenamefont {Kondo}(2009)}]{Ota2009}%
  \BibitemOpen
  \bibfield  {author} {\bibinfo {author} {\bibfnamefont {Yukihiro}\
  \bibnamefont {Ota}}\ and\ \bibinfo {author} {\bibfnamefont {Yasushi}\
  \bibnamefont {Kondo}},\ }\bibfield  {title} {\enquote {\bibinfo {title}
  {Composite pulses in nmr as nonadiabatic geometric quantum gates},}\ }\href
  {\doibase 10.1103/PhysRevA.80.024302} {\bibfield  {journal} {\bibinfo
  {journal} {Phys. Rev. A}\ }\textbf {\bibinfo {volume} {80}},\ \bibinfo
  {pages} {024302} (\bibinfo {year} {2009})}\BibitemShut {NoStop}%
\bibitem [{\citenamefont {Spiegelberg}\ and\ \citenamefont
  {Sj\"oqvist}(2013)}]{Spiegelberg2013}%
  \BibitemOpen
  \bibfield  {author} {\bibinfo {author} {\bibfnamefont {Jakob}\ \bibnamefont
  {Spiegelberg}}\ and\ \bibinfo {author} {\bibfnamefont {Erik}\ \bibnamefont
  {Sj\"oqvist}},\ }\bibfield  {title} {\enquote {\bibinfo {title} {Validity of
  the rotating-wave approximation in nonadiabatic holonomic quantum
  computation},}\ }\href {\doibase 10.1103/PhysRevA.88.054301} {\bibfield
  {journal} {\bibinfo  {journal} {Phys. Rev. A}\ }\textbf {\bibinfo {volume}
  {88}},\ \bibinfo {pages} {054301} (\bibinfo {year} {2013})}\BibitemShut
  {NoStop}%
\bibitem [{\citenamefont {Xu}\ and\ \citenamefont {Long}(2014)}]{Xu2014}%
  \BibitemOpen
  \bibfield  {author} {\bibinfo {author} {\bibfnamefont {Guofu}\ \bibnamefont
  {Xu}}\ and\ \bibinfo {author} {\bibfnamefont {Guilu}\ \bibnamefont {Long}},\
  }\bibfield  {title} {\enquote {\bibinfo {title} {Protecting geometric gates
  by dynamical decoupling},}\ }\href {\doibase 10.1103/PhysRevA.90.022323}
  {\bibfield  {journal} {\bibinfo  {journal} {Phys. Rev. A}\ }\textbf {\bibinfo
  {volume} {90}},\ \bibinfo {pages} {022323} (\bibinfo {year}
  {2014})}\BibitemShut {NoStop}%
\bibitem [{\citenamefont {Liang}\ \emph {et~al.}(2014)\citenamefont {Liang},
  \citenamefont {Du}, \citenamefont {Huang}, \citenamefont {Xue},\ and\
  \citenamefont {Yan}}]{Liang2014}%
  \BibitemOpen
  \bibfield  {author} {\bibinfo {author} {\bibfnamefont {Zhen-Tao}\
  \bibnamefont {Liang}}, \bibinfo {author} {\bibfnamefont {Yan-Xiong}\
  \bibnamefont {Du}}, \bibinfo {author} {\bibfnamefont {Wei}\ \bibnamefont
  {Huang}}, \bibinfo {author} {\bibfnamefont {Zheng-Yuan}\ \bibnamefont {Xue}},
  \ and\ \bibinfo {author} {\bibfnamefont {Hui}\ \bibnamefont {Yan}},\
  }\bibfield  {title} {\enquote {\bibinfo {title} {Nonadiabatic holonomic
  quantum computation in decoherence-free subspaces with trapped ions},}\
  }\href {\doibase 10.1103/PhysRevA.89.062312} {\bibfield  {journal} {\bibinfo
  {journal} {Phys. Rev. A}\ }\textbf {\bibinfo {volume} {89}},\ \bibinfo
  {pages} {062312} (\bibinfo {year} {2014})}\BibitemShut {NoStop}%
\bibitem [{\citenamefont {Xu}\ \emph {et~al.}(2015)\citenamefont {Xu},
  \citenamefont {Liu}, \citenamefont {Zhao},\ and\ \citenamefont
  {Tong}}]{Xu2015}%
  \BibitemOpen
  \bibfield  {author} {\bibinfo {author} {\bibfnamefont {G.~F.}\ \bibnamefont
  {Xu}}, \bibinfo {author} {\bibfnamefont {C.~L.}\ \bibnamefont {Liu}},
  \bibinfo {author} {\bibfnamefont {P.~Z.}\ \bibnamefont {Zhao}}, \ and\
  \bibinfo {author} {\bibfnamefont {D.~M.}\ \bibnamefont {Tong}},\ }\bibfield
  {title} {\enquote {\bibinfo {title} {Nonadiabatic holonomic gates realized by
  a single-shot implementation},}\ }\href {\doibase 10.1103/PhysRevA.92.052302}
  {\bibfield  {journal} {\bibinfo  {journal} {Phys. Rev. A}\ }\textbf {\bibinfo
  {volume} {92}},\ \bibinfo {pages} {052302} (\bibinfo {year}
  {2015})}\BibitemShut {NoStop}%
\bibitem [{\citenamefont {Albert}\ \emph {et~al.}(2016)\citenamefont {Albert},
  \citenamefont {Shu}, \citenamefont {Krastanov}, \citenamefont {Shen},
  \citenamefont {Liu}, \citenamefont {Yang}, \citenamefont {Schoelkopf},
  \citenamefont {Mirrahimi}, \citenamefont {Devoret},\ and\ \citenamefont
  {Jiang}}]{Albert2016}%
  \BibitemOpen
  \bibfield  {author} {\bibinfo {author} {\bibfnamefont {Victor~V.}\
  \bibnamefont {Albert}}, \bibinfo {author} {\bibfnamefont {Chi}\ \bibnamefont
  {Shu}}, \bibinfo {author} {\bibfnamefont {Stefan}\ \bibnamefont {Krastanov}},
  \bibinfo {author} {\bibfnamefont {Chao}\ \bibnamefont {Shen}}, \bibinfo
  {author} {\bibfnamefont {Ren-Bao}\ \bibnamefont {Liu}}, \bibinfo {author}
  {\bibfnamefont {Zhen-Biao}\ \bibnamefont {Yang}}, \bibinfo {author}
  {\bibfnamefont {Robert~J.}\ \bibnamefont {Schoelkopf}}, \bibinfo {author}
  {\bibfnamefont {Mazyar}\ \bibnamefont {Mirrahimi}}, \bibinfo {author}
  {\bibfnamefont {Michel~H.}\ \bibnamefont {Devoret}}, \ and\ \bibinfo {author}
  {\bibfnamefont {Liang}\ \bibnamefont {Jiang}},\ }\bibfield  {title} {\enquote
  {\bibinfo {title} {Holonomic quantum control with continuous variable
  systems},}\ }\href {\doibase 10.1103/PhysRevLett.116.140502} {\bibfield
  {journal} {\bibinfo  {journal} {Phys. Rev. Lett.}\ }\textbf {\bibinfo
  {volume} {116}},\ \bibinfo {pages} {140502} (\bibinfo {year}
  {2016})}\BibitemShut {NoStop}%
\bibitem [{\citenamefont {Zhao}\ \emph {et~al.}(2016)\citenamefont {Zhao},
  \citenamefont {Xu},\ and\ \citenamefont {Tong}}]{Zhao2016}%
  \BibitemOpen
  \bibfield  {author} {\bibinfo {author} {\bibfnamefont {P.~Z.}\ \bibnamefont
  {Zhao}}, \bibinfo {author} {\bibfnamefont {G.~F.}\ \bibnamefont {Xu}}, \ and\
  \bibinfo {author} {\bibfnamefont {D.~M.}\ \bibnamefont {Tong}},\ }\bibfield
  {title} {\enquote {\bibinfo {title} {Nonadiabatic geometric quantum
  computation in decoherence-free subspaces based on unconventional geometric
  phases},}\ }\href {\doibase 10.1103/PhysRevA.94.062327} {\bibfield  {journal}
  {\bibinfo  {journal} {Phys. Rev. A}\ }\textbf {\bibinfo {volume} {94}},\
  \bibinfo {pages} {062327} (\bibinfo {year} {2016})}\BibitemShut {NoStop}%
\bibitem [{\citenamefont {Zhao}\ \emph
  {et~al.}(2017{\natexlab{b}})\citenamefont {Zhao}, \citenamefont {Xu},
  \citenamefont {Ding}, \citenamefont {Sj\"oqvist},\ and\ \citenamefont
  {Tong}}]{Zhao2017a}%
  \BibitemOpen
  \bibfield  {author} {\bibinfo {author} {\bibfnamefont {P.~Z.}\ \bibnamefont
  {Zhao}}, \bibinfo {author} {\bibfnamefont {G.~F.}\ \bibnamefont {Xu}},
  \bibinfo {author} {\bibfnamefont {Q.~M.}\ \bibnamefont {Ding}}, \bibinfo
  {author} {\bibfnamefont {Erik}\ \bibnamefont {Sj\"oqvist}}, \ and\ \bibinfo
  {author} {\bibfnamefont {D.~M.}\ \bibnamefont {Tong}},\ }\bibfield  {title}
  {\enquote {\bibinfo {title} {Single-shot realization of nonadiabatic
  holonomic quantum gates in decoherence-free subspaces},}\ }\href {\doibase
  10.1103/PhysRevA.95.062310} {\bibfield  {journal} {\bibinfo  {journal} {Phys.
  Rev. A}\ }\textbf {\bibinfo {volume} {95}},\ \bibinfo {pages} {062310}
  (\bibinfo {year} {2017}{\natexlab{b}})}\BibitemShut {NoStop}%
\bibitem [{\citenamefont {Zhao}\ \emph {et~al.}(2018)\citenamefont {Zhao},
  \citenamefont {Wu}, \citenamefont {Xing}, \citenamefont {Xu},\ and\
  \citenamefont {Tong}}]{Zhao2018}%
  \BibitemOpen
  \bibfield  {author} {\bibinfo {author} {\bibfnamefont {P.~Z.}\ \bibnamefont
  {Zhao}}, \bibinfo {author} {\bibfnamefont {X.}~\bibnamefont {Wu}}, \bibinfo
  {author} {\bibfnamefont {T.~H.}\ \bibnamefont {Xing}}, \bibinfo {author}
  {\bibfnamefont {G.~F.}\ \bibnamefont {Xu}}, \ and\ \bibinfo {author}
  {\bibfnamefont {D.~M.}\ \bibnamefont {Tong}},\ }\bibfield  {title} {\enquote
  {\bibinfo {title} {Nonadiabatic holonomic quantum computation with rydberg
  superatoms},}\ }\href {\doibase 10.1103/PhysRevA.98.032313} {\bibfield
  {journal} {\bibinfo  {journal} {Phys. Rev. A}\ }\textbf {\bibinfo {volume}
  {98}},\ \bibinfo {pages} {032313} (\bibinfo {year} {2018})}\BibitemShut
  {NoStop}%
\bibitem [{\citenamefont {Zhao}\ \emph
  {et~al.}(2019{\natexlab{a}})\citenamefont {Zhao}, \citenamefont {Xu},\ and\
  \citenamefont {Tong}}]{Zhao2019}%
  \BibitemOpen
  \bibfield  {author} {\bibinfo {author} {\bibfnamefont {P.~Z.}\ \bibnamefont
  {Zhao}}, \bibinfo {author} {\bibfnamefont {G.~F.}\ \bibnamefont {Xu}}, \ and\
  \bibinfo {author} {\bibfnamefont {D.~M.}\ \bibnamefont {Tong}},\ }\bibfield
  {title} {\enquote {\bibinfo {title} {Nonadiabatic holonomic multiqubit
  controlled gates},}\ }\href {\doibase 10.1103/PhysRevA.99.052309} {\bibfield
  {journal} {\bibinfo  {journal} {Phys. Rev. A}\ }\textbf {\bibinfo {volume}
  {99}},\ \bibinfo {pages} {052309} (\bibinfo {year}
  {2019}{\natexlab{a}})}\BibitemShut {NoStop}%
\bibitem [{\citenamefont {Leibfried}\ \emph {et~al.}(2003)\citenamefont
  {Leibfried}, \citenamefont {DeMarco}, \citenamefont {Meyer}, \citenamefont
  {Lucas}, \citenamefont {Barrett}, \citenamefont {Britton}, \citenamefont
  {Itano}, \citenamefont {Jelenković}, \citenamefont {Langer}, \citenamefont
  {Rosenband},\ and\ \citenamefont {Wineland}}]{Leibfried2003}%
  \BibitemOpen
  \bibfield  {author} {\bibinfo {author} {\bibfnamefont {D.}~\bibnamefont
  {Leibfried}}, \bibinfo {author} {\bibfnamefont {B.}~\bibnamefont {DeMarco}},
  \bibinfo {author} {\bibfnamefont {V.}~\bibnamefont {Meyer}}, \bibinfo
  {author} {\bibfnamefont {D.}~\bibnamefont {Lucas}}, \bibinfo {author}
  {\bibfnamefont {M.}~\bibnamefont {Barrett}}, \bibinfo {author} {\bibfnamefont
  {J.}~\bibnamefont {Britton}}, \bibinfo {author} {\bibfnamefont {W.~M.}\
  \bibnamefont {Itano}}, \bibinfo {author} {\bibfnamefont {B.}~\bibnamefont
  {Jelenković}}, \bibinfo {author} {\bibfnamefont {C.}~\bibnamefont {Langer}},
  \bibinfo {author} {\bibfnamefont {T.}~\bibnamefont {Rosenband}}, \ and\
  \bibinfo {author} {\bibfnamefont {D.~J.}\ \bibnamefont {Wineland}},\
  }\bibfield  {title} {\enquote {\bibinfo {title} {Experimental demonstration
  of a robust, high-fidelity geometric two ion-qubit phase gate},}\ }\href
  {https://doi.org/10.1038/nature01492} {\bibfield  {journal} {\bibinfo
  {journal} {Nature}\ }\textbf {\bibinfo {volume} {422}},\ \bibinfo {pages}
  {412--415} (\bibinfo {year} {2003})}\BibitemShut {NoStop}%
\bibitem [{\citenamefont {Du}\ \emph {et~al.}(2006)\citenamefont {Du},
  \citenamefont {Zou},\ and\ \citenamefont {Wang}}]{Du2006}%
  \BibitemOpen
  \bibfield  {author} {\bibinfo {author} {\bibfnamefont {Jiangfeng}\
  \bibnamefont {Du}}, \bibinfo {author} {\bibfnamefont {Ping}\ \bibnamefont
  {Zou}}, \ and\ \bibinfo {author} {\bibfnamefont {Z.~D.}\ \bibnamefont
  {Wang}},\ }\bibfield  {title} {\enquote {\bibinfo {title} {Experimental
  implementation of high-fidelity unconventional geometric quantum gates using
  an nmr interferometer},}\ }\href {\doibase 10.1103/PhysRevA.74.020302}
  {\bibfield  {journal} {\bibinfo  {journal} {Phys. Rev. A}\ }\textbf {\bibinfo
  {volume} {74}},\ \bibinfo {pages} {020302} (\bibinfo {year}
  {2006})}\BibitemShut {NoStop}%
\bibitem [{\citenamefont {Yan}\ \emph {et~al.}(2019)\citenamefont {Yan},
  \citenamefont {Liu}, \citenamefont {Xu}, \citenamefont {Song}, \citenamefont
  {Liu}, \citenamefont {Zhang}, \citenamefont {Deng}, \citenamefont {Yan},
  \citenamefont {Rong}, \citenamefont {Huang}, \citenamefont {Yung},
  \citenamefont {Chen},\ and\ \citenamefont {Yu}}]{Yan2019}%
  \BibitemOpen
  \bibfield  {author} {\bibinfo {author} {\bibfnamefont {Tongxing}\
  \bibnamefont {Yan}}, \bibinfo {author} {\bibfnamefont {Bao-Jie}\ \bibnamefont
  {Liu}}, \bibinfo {author} {\bibfnamefont {Kai}\ \bibnamefont {Xu}}, \bibinfo
  {author} {\bibfnamefont {Chao}\ \bibnamefont {Song}}, \bibinfo {author}
  {\bibfnamefont {Song}\ \bibnamefont {Liu}}, \bibinfo {author} {\bibfnamefont
  {Zhensheng}\ \bibnamefont {Zhang}}, \bibinfo {author} {\bibfnamefont {Hui}\
  \bibnamefont {Deng}}, \bibinfo {author} {\bibfnamefont {Zhiguang}\
  \bibnamefont {Yan}}, \bibinfo {author} {\bibfnamefont {Hao}\ \bibnamefont
  {Rong}}, \bibinfo {author} {\bibfnamefont {Keqiang}\ \bibnamefont {Huang}},
  \bibinfo {author} {\bibfnamefont {Man-Hong}\ \bibnamefont {Yung}}, \bibinfo
  {author} {\bibfnamefont {Yuanzhen}\ \bibnamefont {Chen}}, \ and\ \bibinfo
  {author} {\bibfnamefont {Dapeng}\ \bibnamefont {Yu}},\ }\bibfield  {title}
  {\enquote {\bibinfo {title} {Experimental realization of nonadiabatic
  shortcut to non-abelian geometric gates},}\ }\href {\doibase
  10.1103/PhysRevLett.122.080501} {\bibfield  {journal} {\bibinfo  {journal}
  {Phys. Rev. Lett.}\ }\textbf {\bibinfo {volume} {122}},\ \bibinfo {pages}
  {080501} (\bibinfo {year} {2019})}\BibitemShut {NoStop}%
\bibitem [{\citenamefont {Xu}\ \emph {et~al.}(2020)\citenamefont {Xu},
  \citenamefont {Hua}, \citenamefont {Chen}, \citenamefont {Pan}, \citenamefont
  {Li}, \citenamefont {Han}, \citenamefont {Cai}, \citenamefont {Ma},
  \citenamefont {Wang}, \citenamefont {Song}, \citenamefont {Xue},\ and\
  \citenamefont {Sun}}]{Xu2020}%
  \BibitemOpen
  \bibfield  {author} {\bibinfo {author} {\bibfnamefont {Y.}~\bibnamefont
  {Xu}}, \bibinfo {author} {\bibfnamefont {Z.}~\bibnamefont {Hua}}, \bibinfo
  {author} {\bibfnamefont {Tao}\ \bibnamefont {Chen}}, \bibinfo {author}
  {\bibfnamefont {X.}~\bibnamefont {Pan}}, \bibinfo {author} {\bibfnamefont
  {X.}~\bibnamefont {Li}}, \bibinfo {author} {\bibfnamefont {J.}~\bibnamefont
  {Han}}, \bibinfo {author} {\bibfnamefont {W.}~\bibnamefont {Cai}}, \bibinfo
  {author} {\bibfnamefont {Y.}~\bibnamefont {Ma}}, \bibinfo {author}
  {\bibfnamefont {H.}~\bibnamefont {Wang}}, \bibinfo {author} {\bibfnamefont
  {Y.~P.}\ \bibnamefont {Song}}, \bibinfo {author} {\bibfnamefont {Zheng-Yuan}\
  \bibnamefont {Xue}}, \ and\ \bibinfo {author} {\bibfnamefont
  {L.}~\bibnamefont {Sun}},\ }\bibfield  {title} {\enquote {\bibinfo {title}
  {Experimental implementation of universal nonadiabatic geometric quantum
  gates in a superconducting circuit},}\ }\href {\doibase
  10.1103/PhysRevLett.124.230503} {\bibfield  {journal} {\bibinfo  {journal}
  {Phys. Rev. Lett.}\ }\textbf {\bibinfo {volume} {124}},\ \bibinfo {pages}
  {230503} (\bibinfo {year} {2020})}\BibitemShut {NoStop}%
\bibitem [{\citenamefont {Zhao}\ \emph
  {et~al.}(2019{\natexlab{b}})\citenamefont {Zhao}, \citenamefont {Dong},
  \citenamefont {Zhang}, \citenamefont {Guo}, \citenamefont {Tong},\ and\
  \citenamefont {Yin}}]{zhao2019experimental}%
  \BibitemOpen
  \bibfield  {author} {\bibinfo {author} {\bibfnamefont {P.~Z.}\ \bibnamefont
  {Zhao}}, \bibinfo {author} {\bibfnamefont {Zhangjingzi}\ \bibnamefont
  {Dong}}, \bibinfo {author} {\bibfnamefont {Zhenxing}\ \bibnamefont {Zhang}},
  \bibinfo {author} {\bibfnamefont {Guoping}\ \bibnamefont {Guo}}, \bibinfo
  {author} {\bibfnamefont {D.~M.}\ \bibnamefont {Tong}}, \ and\ \bibinfo
  {author} {\bibfnamefont {Yi}~\bibnamefont {Yin}},\ }\href@noop {} {\enquote
  {\bibinfo {title} {Experimental realization of nonadiabatic geometric gates
  with a superconducting xmon qubit},}\ } (\bibinfo {year}
  {2019}{\natexlab{b}}),\ \Eprint {http://arxiv.org/abs/1909.09970}
  {arXiv:1909.09970 [quant-ph]} \BibitemShut {NoStop}%
\bibitem [{\citenamefont {Haeberlen}\ and\ \citenamefont
  {Waugh}(1968)}]{Haeberlen1968}%
  \BibitemOpen
  \bibfield  {author} {\bibinfo {author} {\bibfnamefont {U.}~\bibnamefont
  {Haeberlen}}\ and\ \bibinfo {author} {\bibfnamefont {J.~S.}\ \bibnamefont
  {Waugh}},\ }\bibfield  {title} {\enquote {\bibinfo {title} {Coherent
  averaging effects in magnetic resonance},}\ }\href {\doibase
  10.1103/PhysRev.175.453} {\bibfield  {journal} {\bibinfo  {journal} {Phys.
  Rev.}\ }\textbf {\bibinfo {volume} {175}},\ \bibinfo {pages} {453--467}
  (\bibinfo {year} {1968})}\BibitemShut {NoStop}%
\bibitem [{\citenamefont {Vandersypen}\ and\ \citenamefont
  {Chuang}(2005)}]{Vandersypen2005}%
  \BibitemOpen
  \bibfield  {author} {\bibinfo {author} {\bibfnamefont {L.~M.~K.}\
  \bibnamefont {Vandersypen}}\ and\ \bibinfo {author} {\bibfnamefont {I.~L.}\
  \bibnamefont {Chuang}},\ }\bibfield  {title} {\enquote {\bibinfo {title} {Nmr
  techniques for quantum control and computation},}\ }\href {\doibase
  10.1103/RevModPhys.76.1037} {\bibfield  {journal} {\bibinfo  {journal} {Rev.
  Mod. Phys.}\ }\textbf {\bibinfo {volume} {76}},\ \bibinfo {pages}
  {1037--1069} (\bibinfo {year} {2005})}\BibitemShut {NoStop}%
\bibitem [{\citenamefont {Goldman}\ and\ \citenamefont
  {Dalibard}(2014)}]{Goldman2014}%
  \BibitemOpen
  \bibfield  {author} {\bibinfo {author} {\bibfnamefont {N.}~\bibnamefont
  {Goldman}}\ and\ \bibinfo {author} {\bibfnamefont {J.}~\bibnamefont
  {Dalibard}},\ }\bibfield  {title} {\enquote {\bibinfo {title} {Periodically
  driven quantum systems: Effective hamiltonians and engineered gauge
  fields},}\ }\href {\doibase 10.1103/PhysRevX.4.031027} {\bibfield  {journal}
  {\bibinfo  {journal} {Phys. Rev. X}\ }\textbf {\bibinfo {volume} {4}},\
  \bibinfo {pages} {031027} (\bibinfo {year} {2014})}\BibitemShut {NoStop}%
\bibitem [{\citenamefont {Bukov}\ \emph {et~al.}(2015)\citenamefont {Bukov},
  \citenamefont {D'Alessio},\ and\ \citenamefont {Polkovnikov}}]{Bukov2015}%
  \BibitemOpen
  \bibfield  {author} {\bibinfo {author} {\bibfnamefont {Marin}\ \bibnamefont
  {Bukov}}, \bibinfo {author} {\bibfnamefont {Luca}\ \bibnamefont {D'Alessio}},
  \ and\ \bibinfo {author} {\bibfnamefont {Anatoli}\ \bibnamefont
  {Polkovnikov}},\ }\bibfield  {title} {\enquote {\bibinfo {title} {Universal
  high-frequency behavior of periodically driven systems: from dynamical
  stabilization to floquet engineering},}\ }\href {\doibase
  10.1080/00018732.2015.1055918} {\bibfield  {journal} {\bibinfo  {journal}
  {Advances in Physics}\ }\textbf {\bibinfo {volume} {64}},\ \bibinfo {pages}
  {139--226} (\bibinfo {year} {2015})}\BibitemShut {NoStop}%
\bibitem [{\citenamefont {Poudel}\ \emph {et~al.}(2015)\citenamefont {Poudel},
  \citenamefont {Ortiz},\ and\ \citenamefont {Viola}}]{Poudel2015}%
  \BibitemOpen
  \bibfield  {author} {\bibinfo {author} {\bibfnamefont {A.}~\bibnamefont
  {Poudel}}, \bibinfo {author} {\bibfnamefont {G.}~\bibnamefont {Ortiz}}, \
  and\ \bibinfo {author} {\bibfnamefont {L.}~\bibnamefont {Viola}},\ }\bibfield
   {title} {\enquote {\bibinfo {title} {Dynamical generation of floquet
  majorana flat bands in s-wave superconductors},}\ }\href
  {http://dx.doi.org/10.1209/0295-5075/110/17004} {\bibfield  {journal}
  {\bibinfo  {journal} {EPL (Europhysics Letters)}\ }\textbf {\bibinfo {volume}
  {110}},\ \bibinfo {pages} {17004} (\bibinfo {year} {2015})}\BibitemShut
  {NoStop}%
\bibitem [{\citenamefont {Eckardt}(2017)}]{Eckardt2017}%
  \BibitemOpen
  \bibfield  {author} {\bibinfo {author} {\bibfnamefont {Andr\'e}\ \bibnamefont
  {Eckardt}},\ }\bibfield  {title} {\enquote {\bibinfo {title} {Colloquium:
  Atomic quantum gases in periodically driven optical lattices},}\ }\href
  {\doibase 10.1103/RevModPhys.89.011004} {\bibfield  {journal} {\bibinfo
  {journal} {Rev. Mod. Phys.}\ }\textbf {\bibinfo {volume} {89}},\ \bibinfo
  {pages} {011004} (\bibinfo {year} {2017})}\BibitemShut {NoStop}%
\bibitem [{\citenamefont {Oka}\ and\ \citenamefont {Kitamura}(2019)}]{Oka2019}%
  \BibitemOpen
  \bibfield  {author} {\bibinfo {author} {\bibfnamefont {Takashi}\ \bibnamefont
  {Oka}}\ and\ \bibinfo {author} {\bibfnamefont {Sota}\ \bibnamefont
  {Kitamura}},\ }\bibfield  {title} {\enquote {\bibinfo {title} {Floquet
  engineering of quantum materials},}\ }\href {\doibase
  10.1146/annurev-conmatphys-031218-013423} {\bibfield  {journal} {\bibinfo
  {journal} {Annual Review of Condensed Matter Physics}\ }\textbf {\bibinfo
  {volume} {10}},\ \bibinfo {pages} {387--408} (\bibinfo {year}
  {2019})}\BibitemShut {NoStop}%
\bibitem [{\citenamefont {Novi\ifmmode~\check{c}\else \v{c}\fi{}enko}\ \emph
  {et~al.}(2017)\citenamefont {Novi\ifmmode~\check{c}\else \v{c}\fi{}enko},
  \citenamefont {Anisimovas},\ and\ \citenamefont {Juzeli\ifmmode~\bar{u}\else
  \={u}\fi{}nas}}]{Novifmmodeheckclsecienko2017}%
  \BibitemOpen
  \bibfield  {author} {\bibinfo {author} {\bibfnamefont {Viktor}\ \bibnamefont
  {Novi\ifmmode~\check{c}\else \v{c}\fi{}enko}}, \bibinfo {author}
  {\bibfnamefont {Egidijus}\ \bibnamefont {Anisimovas}}, \ and\ \bibinfo
  {author} {\bibfnamefont {Gediminas}\ \bibnamefont
  {Juzeli\ifmmode~\bar{u}\else \={u}\fi{}nas}},\ }\bibfield  {title} {\enquote
  {\bibinfo {title} {Floquet analysis of a quantum system with modulated
  periodic driving},}\ }\href {\doibase 10.1103/PhysRevA.95.023615} {\bibfield
  {journal} {\bibinfo  {journal} {Phys. Rev. A}\ }\textbf {\bibinfo {volume}
  {95}},\ \bibinfo {pages} {023615} (\bibinfo {year} {2017})}\BibitemShut
  {NoStop}%
\bibitem [{\citenamefont {Novi\ifmmode~\check{c}\else \v{c}\fi{}enko}\ and\
  \citenamefont {Juzeli\ifmmode~\bar{u}\else
  \={u}\fi{}nas}(2019)}]{PhysRevA.100.012127}%
  \BibitemOpen
  \bibfield  {author} {\bibinfo {author} {\bibfnamefont {Viktor}\ \bibnamefont
  {Novi\ifmmode~\check{c}\else \v{c}\fi{}enko}}\ and\ \bibinfo {author}
  {\bibfnamefont {Gediminas}\ \bibnamefont {Juzeli\ifmmode~\bar{u}\else
  \={u}\fi{}nas}},\ }\bibfield  {title} {\enquote {\bibinfo {title}
  {Non-abelian geometric phases in periodically driven systems},}\ }\href
  {\doibase 10.1103/PhysRevA.100.012127} {\bibfield  {journal} {\bibinfo
  {journal} {Phys. Rev. A}\ }\textbf {\bibinfo {volume} {100}},\ \bibinfo
  {pages} {012127} (\bibinfo {year} {2019})}\BibitemShut {NoStop}%
\bibitem [{\citenamefont {Bomantara}\ and\ \citenamefont
  {Gong}(2018{\natexlab{a}})}]{Bomantara2018}%
  \BibitemOpen
  \bibfield  {author} {\bibinfo {author} {\bibfnamefont {Raditya~Weda}\
  \bibnamefont {Bomantara}}\ and\ \bibinfo {author} {\bibfnamefont {Jiangbin}\
  \bibnamefont {Gong}},\ }\bibfield  {title} {\enquote {\bibinfo {title}
  {Quantum computation via floquet topological edge modes},}\ }\href {\doibase
  10.1103/PhysRevB.98.165421} {\bibfield  {journal} {\bibinfo  {journal} {Phys.
  Rev. B}\ }\textbf {\bibinfo {volume} {98}},\ \bibinfo {pages} {165421}
  (\bibinfo {year} {2018}{\natexlab{a}})}\BibitemShut {NoStop}%
\bibitem [{\citenamefont {Bomantara}\ and\ \citenamefont
  {Gong}(2018{\natexlab{b}})}]{Bomantara2018a}%
  \BibitemOpen
  \bibfield  {author} {\bibinfo {author} {\bibfnamefont {Raditya~Weda}\
  \bibnamefont {Bomantara}}\ and\ \bibinfo {author} {\bibfnamefont {Jiangbin}\
  \bibnamefont {Gong}},\ }\bibfield  {title} {\enquote {\bibinfo {title}
  {Simulation of non-abelian braiding in majorana time crystals},}\ }\href
  {\doibase 10.1103/PhysRevLett.120.230405} {\bibfield  {journal} {\bibinfo
  {journal} {Phys. Rev. Lett.}\ }\textbf {\bibinfo {volume} {120}},\ \bibinfo
  {pages} {230405} (\bibinfo {year} {2018}{\natexlab{b}})}\BibitemShut
  {NoStop}%
\bibitem [{\citenamefont {Liu}\ \emph {et~al.}(2020{\natexlab{b}})\citenamefont
  {Liu}, \citenamefont {Xue},\ and\ \citenamefont {Yung}}]{Liu2020a}%
  \BibitemOpen
  \bibfield  {author} {\bibinfo {author} {\bibfnamefont {Bao-Jie}\ \bibnamefont
  {Liu}}, \bibinfo {author} {\bibfnamefont {Zheng-Yuan}\ \bibnamefont {Xue}}, \
  and\ \bibinfo {author} {\bibfnamefont {Man-Hong}\ \bibnamefont {Yung}},\
  }\href@noop {} {\enquote {\bibinfo {title} {Brachistochronic non-adiabatic
  holonomic quantum control},}\ } (\bibinfo {year} {2020}{\natexlab{b}}),\
  \Eprint {http://arxiv.org/abs/2001.05182} {arXiv:2001.05182 [quant-ph]}
  \BibitemShut {NoStop}%
\bibitem [{\citenamefont {Lloyd}(1995)}]{Lloyd1995}%
  \BibitemOpen
  \bibfield  {author} {\bibinfo {author} {\bibfnamefont {Seth}\ \bibnamefont
  {Lloyd}},\ }\bibfield  {title} {\enquote {\bibinfo {title} {Almost any
  quantum logic gate is universal},}\ }\href {\doibase
  10.1103/PhysRevLett.75.346} {\bibfield  {journal} {\bibinfo  {journal} {Phys.
  Rev. Lett.}\ }\textbf {\bibinfo {volume} {75}},\ \bibinfo {pages} {346--349}
  (\bibinfo {year} {1995})}\BibitemShut {NoStop}%
\bibitem [{\citenamefont {Saffman}\ \emph {et~al.}(2010)\citenamefont
  {Saffman}, \citenamefont {Walker},\ and\ \citenamefont
  {M\o{}lmer}}]{Saffman2010}%
  \BibitemOpen
  \bibfield  {author} {\bibinfo {author} {\bibfnamefont {M.}~\bibnamefont
  {Saffman}}, \bibinfo {author} {\bibfnamefont {T.~G.}\ \bibnamefont {Walker}},
  \ and\ \bibinfo {author} {\bibfnamefont {K.}~\bibnamefont {M\o{}lmer}},\
  }\bibfield  {title} {\enquote {\bibinfo {title} {Quantum information with
  rydberg atoms},}\ }\href {\doibase 10.1103/RevModPhys.82.2313} {\bibfield
  {journal} {\bibinfo  {journal} {Rev. Mod. Phys.}\ }\textbf {\bibinfo {volume}
  {82}},\ \bibinfo {pages} {2313--2363} (\bibinfo {year} {2010})}\BibitemShut
  {NoStop}%
\bibitem [{\citenamefont {Saffman}(2016)}]{Saffman2016}%
  \BibitemOpen
  \bibfield  {author} {\bibinfo {author} {\bibfnamefont {M.}~\bibnamefont
  {Saffman}},\ }\bibfield  {title} {\enquote {\bibinfo {title} {Quantum
  computing with atomic qubits and rydberg interactions: progress and
  challenges},}\ }\href {http://dx.doi.org/10.1088/0953-4075/49/20/202001}
  {\bibfield  {journal} {\bibinfo  {journal} {Journal of Physics B: Atomic,
  Molecular and Optical Physics}\ }\textbf {\bibinfo {volume} {49}},\ \bibinfo
  {pages} {202001} (\bibinfo {year} {2016})}\BibitemShut {NoStop}%
\bibitem [{\citenamefont {Omran}\ \emph {et~al.}(2019)\citenamefont {Omran},
  \citenamefont {Levine}, \citenamefont {Keesling}, \citenamefont {Semeghini},
  \citenamefont {Wang}, \citenamefont {Ebadi}, \citenamefont {Bernien},
  \citenamefont {Zibrov}, \citenamefont {Pichler}, \citenamefont {Choi},
  \citenamefont {Cui}, \citenamefont {Rossignolo}, \citenamefont {Rembold},
  \citenamefont {Montangero}, \citenamefont {Calarco}, \citenamefont {Endres},
  \citenamefont {Greiner}, \citenamefont {Vuleti{\'c}},\ and\ \citenamefont
  {Lukin}}]{omran570}%
  \BibitemOpen
  \bibfield  {author} {\bibinfo {author} {\bibfnamefont {A.}~\bibnamefont
  {Omran}}, \bibinfo {author} {\bibfnamefont {H.}~\bibnamefont {Levine}},
  \bibinfo {author} {\bibfnamefont {A.}~\bibnamefont {Keesling}}, \bibinfo
  {author} {\bibfnamefont {G.}~\bibnamefont {Semeghini}}, \bibinfo {author}
  {\bibfnamefont {T.~T.}\ \bibnamefont {Wang}}, \bibinfo {author}
  {\bibfnamefont {S.}~\bibnamefont {Ebadi}}, \bibinfo {author} {\bibfnamefont
  {H.}~\bibnamefont {Bernien}}, \bibinfo {author} {\bibfnamefont {A.~S.}\
  \bibnamefont {Zibrov}}, \bibinfo {author} {\bibfnamefont {H.}~\bibnamefont
  {Pichler}}, \bibinfo {author} {\bibfnamefont {S.}~\bibnamefont {Choi}},
  \bibinfo {author} {\bibfnamefont {J.}~\bibnamefont {Cui}}, \bibinfo {author}
  {\bibfnamefont {M.}~\bibnamefont {Rossignolo}}, \bibinfo {author}
  {\bibfnamefont {P.}~\bibnamefont {Rembold}}, \bibinfo {author} {\bibfnamefont
  {S.}~\bibnamefont {Montangero}}, \bibinfo {author} {\bibfnamefont
  {T.}~\bibnamefont {Calarco}}, \bibinfo {author} {\bibfnamefont
  {M.}~\bibnamefont {Endres}}, \bibinfo {author} {\bibfnamefont
  {M.}~\bibnamefont {Greiner}}, \bibinfo {author} {\bibfnamefont
  {V.}~\bibnamefont {Vuleti{\'c}}}, \ and\ \bibinfo {author} {\bibfnamefont
  {M.~D.}\ \bibnamefont {Lukin}},\ }\bibfield  {title} {\enquote {\bibinfo
  {title} {Generation and manipulation of schr{\"o}dinger cat states in rydberg
  atom arrays},}\ }\href {\doibase 10.1126/science.aax9743} {\bibfield
  {journal} {\bibinfo  {journal} {Science}\ }\textbf {\bibinfo {volume}
  {365}},\ \bibinfo {pages} {570--574} (\bibinfo {year} {2019})}\BibitemShut
  {NoStop}%
\end{thebibliography}%

\end{document}